\documentclass[usenatbib,useAMS]{mn2e}
\usepackage{graphicx}
\usepackage{aas_macros}
\usepackage{amssymb}
\usepackage[pdftex]{hyperref}

\newcommand{\ergs}{erg s$^{-1}$}
\newcommand{\mics}{$\mu$m}
\newcommand{\msun}{M$_{\odot}$}

\newcommand{\lrad}{$L_{\rm 144}$}
\newcommand{\lsix}{$L_{\rm 6 \mu m}$}

\title[Red QSOs in LoTSS]{Fundamental differences in the radio properties of red and blue quasars: Insight from the LOFAR Two-metre Sky Survey (LoTSS)}

\author[D. J. Rosario et al.]
{D. J.~Rosario$^{1}$,
V. A.~Fawcett$^{1}$,
L. Klindt$^{1}$,
D. M.~Alexander$^{1}$,
L. K.~Morabito$^{1}$,
\newauthor 
S.~Fotopoulou$^{2}$,
E.~Lusso$^{3,4}$,
and G.~Calistro Rivera$^{5}$
\\
$^{1}$Centre for Extragalactic Astronomy, Department of Physics, Durham University, South Road, DH1 3LE, Durham, UK \\
$^{2}$School of Physics, University of Bristol, UK \\
$^{3}$Dipartimento di Fisica e Astronomia, Universit\`a di Firenze, via G. Sansone 1, 50019 Sesto Fiorentino, Firenze, Italy \\
$^{4}$INAF -- Osservatorio Astrofisico di Arcetri, Largo Enrico Fermi 5, 50125 Firenze, Italy\\
$^{5}$European Southern Observatory,  Karl-Schwarzschild-Strasse 2, 85748, Garching bei München, Germany
}

\hypersetup{draft}
\begin{document}

\maketitle

\begin{abstract}

Red quasi-stellar objects (QSOs) are a subset of the luminous end of the cosmic population of active
galactic nuclei (AGN), most of which are reddened by intervening dust along the line-of-sight towards their 
central engines. In recent work from our team, we developed a systematic technique to select red QSOs 
from the Sloan Digital Sky Survey (SDSS), and demonstrated that they have distinctive radio properties using the 
Faint Images of the Radio Sky at Twenty centimeters (FIRST) radio survey. Here we
expand our study using low-frequency radio data from the LOFAR Two-metre Sky Survey (LoTSS). With 
the improvement in depth that LoTSS offers, we confirm key results: 
compared to a control sample of normal ``blue'' QSOs matched in redshift and accretion power, 
red QSOs have a higher radio detection rate and a higher incidence of compact radio morphologies. For the first time,
we also demonstrate that these differences arise primarily in sources of intermediate radio-loudness: 
radio-intermediate red QSOs are $\times 3$ more common than typical QSOs, but the 
excess diminishes among the most radio-loud and the most radio-quiet
systems in our study. We develop Monte-Carlo simulations to explore whether differences in star formation 
could explain these results, and conclude that, while star formation is an important source of low-frequency emission among radio-quiet QSOs,
a population of AGN-driven compact radio sources is the most likely cause for the distinct low-frequency radio properties of red QSOs. 
Our study substantiates the conclusion that fundamental differences must exist between the red and normal blue 
QSO populations.

\end{abstract}

\section{Introduction}

Quasi-stellar objects (QSOs; often also called quasars) are powerful Type 1 Active Galactic Nuclei (AGN) that
mark the sites of the most substantial accretion-powered growth of 
supermassive black holes in the Universe. 
They can be identified by their extreme luminosities and 
unique spectral features \citep{croom04, schneider10} even in the early Universe \citep[e.g.,][]{fan01}.

Most QSOs display blue ultraviolet-optical spectral slopes, consistent with effective
temperatures of their inner accretion discs of $\sim 10^{5}$ K. This allows the efficient identification
of QSOs from multi-band imaging \citep[e.g.,][]{sandage65, richards02}, 
making them an optimal target population for large spectroscopic surveys such as the Sloan Digital Sky Survey (SDSS).
While most QSOs have blue colours, a minority \citep[$\approx  10$\%;][]{richards03,klindt19} have red optical colours
indicative of mild or moderate dust reddening along the line of sight (E(B-V) $\gtrsim 0.04$). 
These red QSOs likely represent the detectable end of a population of more heavily extinguished
luminous AGN, some of which have been uncovered through large-area
near-infrared surveys \citep{glikman07, banerji12}. 

Within the scope of the classical unified model of AGN, in which the accretion disc and broad-line region are
obscured by a canonical obscuring structure, the nuclear dusty `torus' \citep[][and references therein]{netzer15}, 
red QSOs are simply normal blue QSOs that are viewed at grazing incidence to the torus. 
However, this view has been challenged by numerous
lines of evidence which demonstrate that red QSOs have distinctive properties that cannot be explained by
differences in torus orientation. Compared to typical QSOs, their bolometric luminosity functions are flatter \citep{banerji15}, 
their host galaxies are more often in major mergers \citep{urrutia08, glikman15}, and they may show a higher incidence
of strong AGN outflows \citep{urrutia09}.

The classical radio bands offer some unique advantages towards the study of QSOs, particularly in tests of AGN unification.
Most radio emission from galaxies arises from optically thin synchrotron-emitting plasma, 
and is not substantially absorbed by intervening dust or cold gas.
In the context of AGN, radio emission is therefore an excellent orientation-insensitive tracer of the nuclear output.
Indeed, AGN dominate the bright radio source population that comprise the majority of large-area or all-sky surveys
\citep[e.g.,][]{helfand15}. At fainter fluxes, non-AGN processes, such as star formation (SF), contribute
significantly to the radio emission of detectable sources \citep[][and references therein]{panessa19}.

Early work has suggested a link between the occurrence of radio emission and the optical colours of QSOs 
\citep{ivezic02,richards03,white03,white07,tsai17}, but until recently, there lacked a detailed and systematic study
which controlled for important systematics such as survey selection effects, AGN luminosity, and cosmic
evolution of the QSO population. Harnessing the extensive spectroscopic sample and 
multi-wavelength constraints available over the SDSS, our team has devised a redshift-insensitive approach to select red QSOs
which mitigates these biases. Using the SDSS DR7 QSO catalogue and the FIRST 1.4 GHz radio survey, our foundational
study identified a number of essential radio properties that are preferentially found among red QSOs \citep{klindt19}.
These systems show a detection rate in FIRST that is up to three times higher than normal QSOs of 
equivalent luminosity at the same redshifts.
The radio sources responsible for the enhancement are inevitably compact, showing no extended
structure beyond the FIRST imaging beam of $5$''. They are also often faint in the radio; the most
substantial enhancement in the radio source population among red QSOs occurs among ``radio-quiet''
sources (see \citet{klindt19} or Section \ref{R_results} for a definition of radio-loudness). 

A number of large-area modern radio surveys are currently in operation, giving us the opportunity to validate
and expand upon the key work of \citet{klindt19}. Especially interesting in this regard are the next-generation
low-frequency radio surveys from the International Low-Frequency Array (LOFAR), particularly the wide-area component of the 
LOFAR Two-metre Sky Survey (LoTSS), which will eventually cover a contiguous area of 20,000 deg$^{2}$. 
For sources with typical synchrotron spectra, LoTSS is an order of magnitude more sensitive than FIRST
(Section \ref{det_stats}). In addition,
the spectral baseline available between the observing bands of LoTSS ($\approx 144$ MHz) and FIRST ($\approx 1.4$ GHz) 
yield additional useful diagnostics into the radio properties of QSOs. LoTSS also
allows us to ameliorate potential biases related to the fact that a fraction of the SDSS QSOs were themselves
selected for spectroscopic follow-up using the FIRST radio survey \citep{richards02}.

In this work, we take advantage of the expanded, deeper, and more complete database of spectroscopic 
QSOs from the SDSS DR14 QSO catalogue \citep{paris18}, combining it with low-frequency radio
imaging and photometry from the LoTSS first full-resolution data release (DR1). Section \ref{data}
presents the datasets and the methodology of our analysis.
Taking an approach that parallels \citet{klindt19}, we contrast the radio detection rates (Section \ref{det_stats}), 
luminosities and spectral indices (Section \ref{lradio_dists}), radio-loudness (Section \ref{R_results}) and 
morphologies (Section \ref{radio_morphs}) of both red and normal QSOs. 
In Section \ref{discussion}, we explore the origin of the radio emission in QSOs and 
conclude this work by folding together our results into a discussion of the likely origin of the unique 
radio properties of red QSOs.

Throughout this work, we assume a concordance cosmology \citep{planck16} and implement it in our calculations 
using software built into the AstroPy \texttt{cosmology} module. 

\section{Data and Methods} \label{data}

\subsection{SDSS colour-selected QSOs} \label{sample_description}

\begin{figure}
\centering 
\includegraphics[width=\columnwidth]{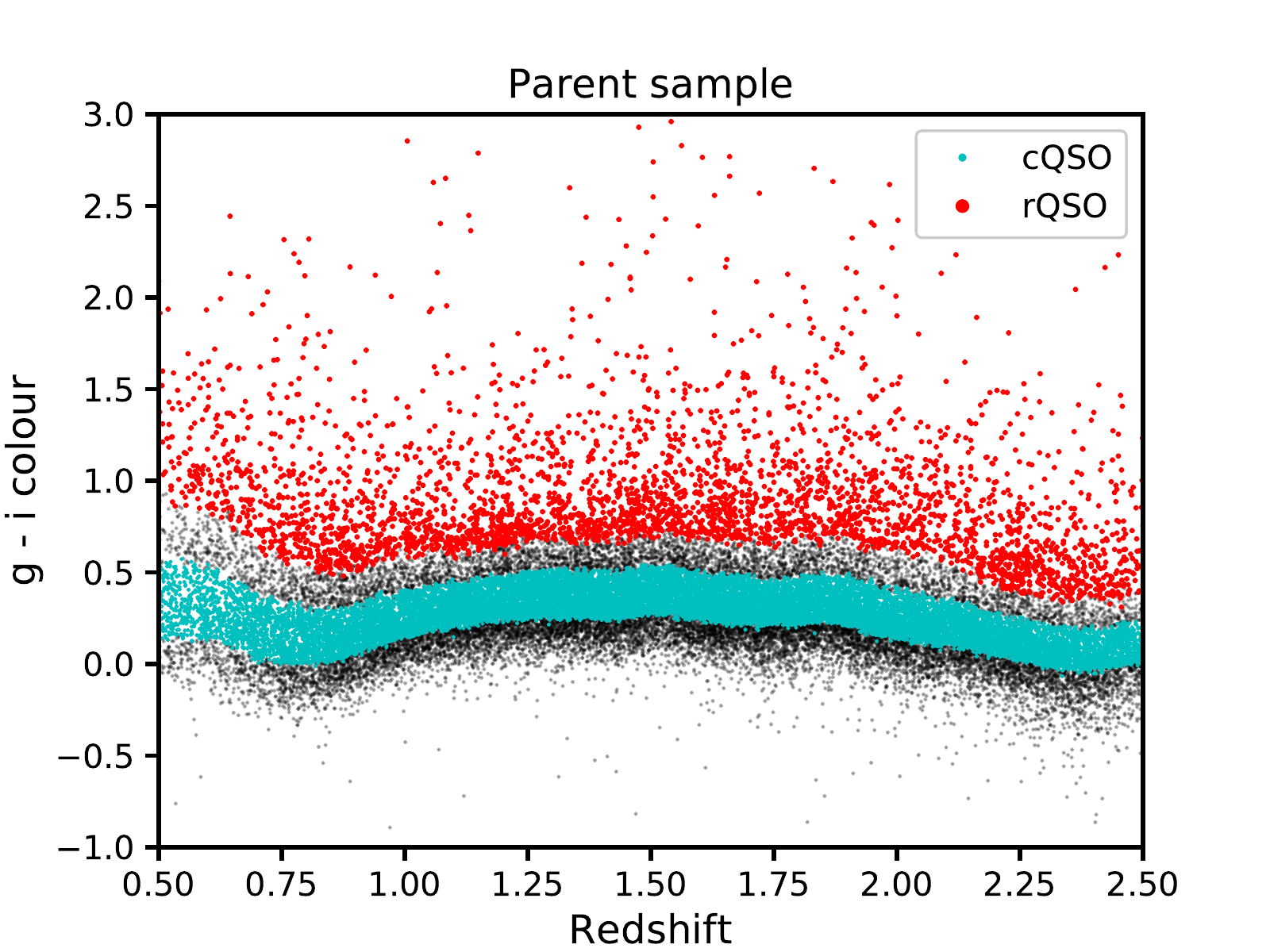}
\caption
{Observed $g-i$ colour plotted against redshift for the $\approx 43000$ SDSS DR14 QSOs that lie in the region covered by the LoTSS DR1
field (see Section \ref{lotss_data}). Only QSOs in the redshift range of $0.5<z<2.5$ are shown.
Red QSOs (rQSOs) and control QSOs (cQSOs) from the `parent sample' 
(see Section \ref{sample_description}) are shown as red and cyan points. These constitute the `colour-selected' QSOs used
in this work, while the black points indicate other QSOs excluded from our analysis. Note that the range of colours used
to define these samples is redshift-dependent as the observed $g$ and $i$ bands sample different parts of a typical QSO
spectral energy distribution (SED) with redshift. rQSOs span a much larger range in colour than cQSOs.
}
\label{gicolz}
\end{figure}

The most recent edition of the SDSS QSO catalogue \citep[DR14;][]{paris18} contains more than 500K entries from a number
of different QSO spectroscopic surveys under the SDSS umbrella, including those from the classical SDSS I/II surveys,
BOSS, e-BOSS, and dedicated follow-up programs such as SPIDERS \citep[][and references therein]{paris18}.
In addition to spectroscopic redshifts, the catalogue includes photometry in the five SDSS bands ($ugriz$) and the
amount of foreground Galactic extinction in each band. When SDSS fluxes and colours are discussed in this work, the reader
should assume, unless otherwise stated, that they have been corrected for Galactic extinction following the 
recommendations of \citet{paris18}.

We selected red QSOs (hereafter, rQSOs) following the approach described in \citet{klindt19} and illustrated in Figure \ref{gicolz}.
For our selection, we considered all SDSS DR14 QSOs in the redshift range $0.5<z<2.5$, motivated by the desire to exclude 
lower luminosity Seyfert 1 galaxies which populate the SDSS DR14 catalogue at low redshifts, and to avoid the 
effects of the Lyman break on our colour selection at $z>2.5$.
Working in contiguous, equally-populated redshift bins of 1000 objects each, we
selected rQSOs as the 10\% of population with the reddest $g-i$ colours. This process identifies a set
of rQSOs that is unbiased by the systematic observed-frame colour variation of the SDSS QSO population with redshift.
From the same binned samples, we also selected the 50\% of QSOs with colours
that spanned the median $g-i$ colour (hereafter, cQSOs). These systems represent the typical ``blue'' QSO population and
serve as a set of controls for much of our further analysis. 

From the 43124 SDSS DR14 QSOs in the region of sky with LoTSS coverage (see Section \ref{lotss_data}), 
we arrive at a set of 3568 rQSOs and 17316 cQSOs, which we refer to as the `parent sample' in the rest of this work.
Figure \ref{gicolz} summarises the effect of our observed-frame colour-based selection as it pertains
to these QSOs.

While broadly based on the approach of \citet{klindt19}, our selection is different in detail. \citet{klindt19} identify
QSOs from the SDSS DR7 `uniform' sample, a set of objects subject to magnitude and colour cuts. Their work determined
that the restriction to a uniform sample did not strongly influence their results, and therefore, we relax this criterion.
In addition, \citet{klindt19} identified a control sample as the 10\% of QSOs around the median redshift-dependent colour, while we use a
substantially more relaxed 50\% cQSO cut. This again relies on the assertion from \citet{klindt19} that the radio properties were only
deviant among the reddest of the SDSS QSO population. Our more liberal cQSO selection approach provides us with a larger control
sample allowing for more detailed statistical tests. An accompanying study by \citet{fawcett20}, which examines the radio properties of
SDSS QSOs in the SDSS Stripe82 and the COSMOS fields, also employs a similar colour-based selection strategy as in this paper.


\subsection{{\it WISE} photometry} \label{wise_data}

\begin{figure*}
\centering 
\includegraphics[width=\textwidth]{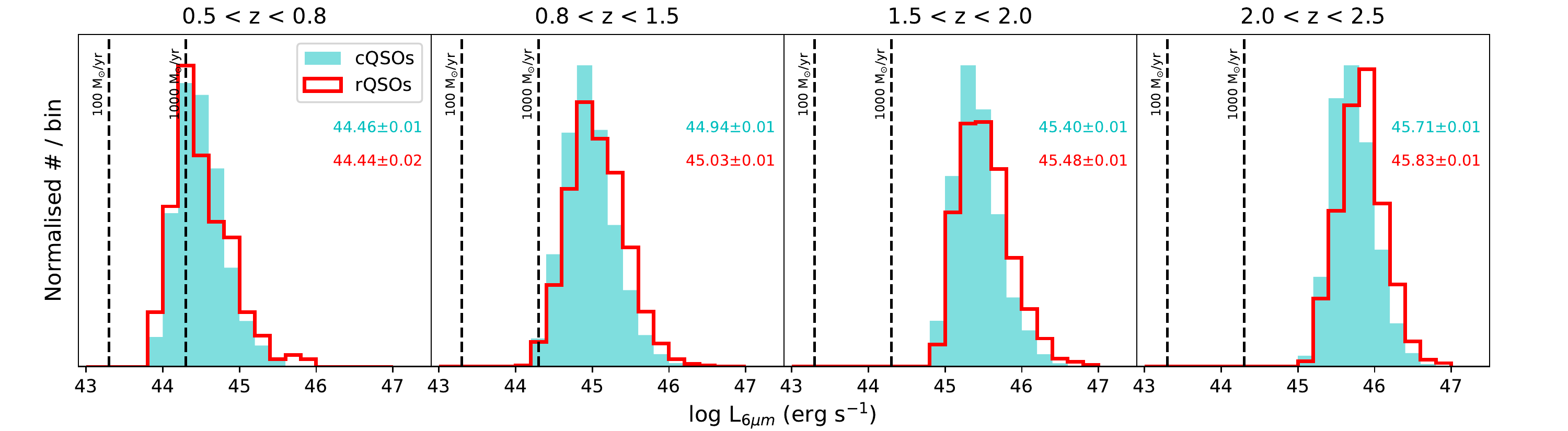}
\caption
{The distributions of rest-frame 6 \mics\ luminosity (\lsix) of cQSOs (cyan) and rQSOs (red)
from the working sample (defined in Section \ref{sample_description}) in four redshift bins.
All distributions have been normalised to unity for ease of visual comparison.
The median \lsix\ and its error for cQSOs (cyan text) and rQSOs (red text) are written within each panel.
The dashed vertical lines give the 6 \mics\ luminosities of powerful starbursts with the indicated SFRs,
scaled from the local SF-dominated ULIRG Arp 220.
In our lowest redshift bin, a non-negligible fraction of \lsix\ from the lowest-luminosity QSOs could potentially
arise from strong SF rather than AGN-heated dust. 
At higher redshifts, our sample consists of more luminous AGN that are too
powerful to be contaminated by SF in the MIR.
}
\label{l6comp}
\end{figure*}

Rest-frame mid-infrared (MIR) emission in the vast majority of SDSS QSOs is dominated by AGN-heated dust \citep[e.g.][]{richards06}, 
which can serve as a useful extinction-free measure of the intrinsic power of the AGN\footnote{Based on the \citet{draine03}
IR interstellar extinction law, the MIR extinction expected from the dust column inferred among rQSOs is $< 0.05$ mag, essentially
negligible for the purposes of this work.}.
This property is particularly important 
when comparing cQSOs and rQSOs: without the application of poorly-constrained extinction corrections
the additional dust extinction in red QSOs leads to depressed estimates of their UV/optical-based AGN luminosities \citep{klindt19}. 

We use photometry from the {\it WISE} all-sky survey to estimate the rest-frame 6 \mics\ luminosity (\lsix) of the QSOs.
To this end, we imposed on our sample a S/N$>2$ detection in the three shorter {\it WISE} bands (W1 [3.5\mics], W2 [4.6 \mics], and W3 [12 \mics]). We compiled this photometry from the ALLWISE catalogue available on IRSA, using a cone search with a tolerance of $2\farcs7$, which
ensures a false-match probability of $<1$\% \citep{lake12, klindt19}. 

\lsix\ was calculated using W2 and W3 fluxes, expressed per logarithmic flux interval (i.e., $\nu$f$_{\nu}$) by multiplying the
flux densities by the pivot frequencies of the respective filter profiles. These fluxes were interpolated/extrapolated in log-linear fashion 
to the redshift-dependent observed-frame wavelength corresponding to rest-frame 6 \mics. The resultant flux was converted to a luminosity 
using the cosmological luminosity distance. While we did not explicitly use the W1 flux in these calculations, we require a detection in this
band to ensure clean counterpart associations with the LoTSS radio sources (see Section \ref{lotss_data}).
About 56\% of QSOs from the parent sample have the requisite photometry in the three {\it WISE} bands. Henceforth,
we refer to the DR14 colour-selected QSOs in the redshift range $0.5<z<2.5$ with \lsix\ estimates
(i.e., a S/N$>2$ detection in W1, W2 \& W3) as the `working sample': a total of 2542 rQSOs and 9212 cQSOs.

Our working sample covers a large range in redshift. To minimise the confounding effects of Malmquist bias,
we divide the QSOs in the working sample into four redshift bins for most of our analysis: $0.5 \leq z < 0.8$,
$0.8 \leq z < 1.5$, $1.5 \leq z < 2.0$, and $2.0 \leq z < 2.5$.

For the rest of this work, we assume that the 6 \mics\ bolometric corrections of cQSOs and rQSOs are indistinguishable, i.e.,
an identical average fraction of the intrinsic luminosities of their AGNs are reprocessed by their dusty tori. For this to be the case,
rQSOs and cQSOs must have the same distribution of torus covering factors. If the structures of the tori were very different
between the two colour-selected QSO subsets, one might expect them to display systematic differences in their MIR colours,
since the relative proportion of dust of different temperatures is sensitive to torus geometry.
However, \citet{klindt19} demonstrated that the {\it WISE} colours of the different QSO subsets are very similar, 
supporting our working assumption. Future work from our team aims to robustly test this notion using
detailed spectroscopic and SED modelling, but for now, we proceed with the assumption that the
cQSOs and rQSOs have intrinsically similar tori, and therefore, similar 6 \mics\ bolometric corrections.
 
Figure \ref{l6comp} compares the \lsix\ distributions of cQSOs and rQSOs from the working sample in our
four main redshift bins. In the three higher redshift bins ($z>0.8$), the median \lsix\ of the rQSOs is
higher than the cQSOs with high statistical significance: two-sided Kolmogorov-Smirnov (KS) tests indicate that the respective
distributions are inconsistent at a probability of $>99$\%. 
As discussed in \citet{klindt19}, this can be understood as a consequence of the optical 
incompleteness suffered by red QSOs in the SDSS. In the rest-frame optical/UV, rQSOs are fainter than equally powerful cQSOs
due to their higher dust extinction. Therefore, the rQSOs that are optically bright enough for spectroscopic targeting and 
characterisation by the SDSS will generally be more luminous than cQSOs when using an extinction-insensitive measure
such as \lsix.

At $0.5 \leq z < 0.8$, rQSOs and cQSOs are equally luminous.
Among the lower luminosity QSOs at these redshifts, emission from dust heated by 
strong starbursts can significantly contaminate \lsix.
The vertical dashed lines in Figure \ref{l6comp} show a characteristic MIR luminosity from extreme starbursts with different 
star formation rates (SFRs). If a significant fraction of rQSOs at low redshifts are reddened not by dust, but 
through the contamination of their optical and MIR light from a powerful on-going starburst, this may explain why their 
MIR luminosities are similar to normal bright QSOs with blue colours. This is consistent with the conclusions of
\citet{klindt19} that, at lower redshifts, there exists a subpopulation in which a lower luminosity Type 1 AGN in a 
massive host galaxy satisfies our simple colour-based selection of rQSOs. Towards higher redshifts, the nuclear
luminosities involved are too high to suffer significantly from this effect, leading to a much purer dust-reddened
rQSO sample.

\subsection{LoTSS DR1} \label{lotss_data}

The LOFAR Two-metre Sky Survey is an on-going programme to image the entire northern sky in the $120$--$168$ MHz radio band
at a resolution of $6$'' \citep{shimwell17}. The first full resolution data release \citep[DR1;][]{shimwell19} covers 424
deg$^{2}$ in the HETDEX spring field (RA: 10h45m -- 15h30m, Dec: 45$^{\circ}$ -- 57$^{\circ}$), which lies completely within
the SDSS footprint. Calibrated and synthesised radio images, in the form of science mosaics and root-mean-square (RMS) noise maps, 
and a combined source catalogue are available from the 
\href{http://lofar-surveys.org/releases.html}{LOFAR surveys public releases} page.

We selected SDSS DR14 QSOs in the LoTSS DR1 area by identifying those that have a measured value in the RMS maps 
at their sky positions. 
This approach is superior to a simple footprint-based selection, as it allows us to account for the patchy,  
non-uniform sky coverage of the radio survey.
While the LoTSS radio imaging has a median sensitivity of 71 $\mu$Jy  beam$^{-1}$, the noise varies considerably over the field, 
and is particularly high around bright sources due to side lobe residuals that remain after deconvolution of the maps \citep{shimwell19}.
We perform the majority of the analysis in this paper -- radio detection statistics and morphological analysis -- on all QSOs 
in the LoTSS DR1 area, irrespective of the local noise level. However, we have checked that our primary results hold after restricting
the sample to those in areas of low noise. While this restriction is better for uniform statistics, it biases our study against those QSOs
with the brightest radio counterparts since these sources alter the local RMS in their regions of the maps.

The LoTSS images do not have a uniform point-spread function (PSF) due to ionospheric variations, astrometric errors, 
the deconvolution masking strategy, and time- and frequency-dependent station beams and smearing. 
As a consequence, it is not trivial to determine whether a source is unresolved or not, and 
occasionally absolute astrometric errors can be $>1$''.

To help ameliorate positional uncertainties and optimise the use of the radio survey for multi-wavelength work, 
the LoTSS DR1 release includes a catalogue of associations to optical and infrared (IR) counterparts \citep{williams19}.
This catalogue is the product of likelihood-ratio matching using colour and magnitude information, 
as well as the visual examination of 13,000 extended or complex sources, to identify the best {\it WISE} W1 and/or PanSTARRS 
counterparts to each radio source. Multiple components for some large radio sources were reduced to single entries, integrating the fluxes
of the components and tying the central or core component with the best optical/IR counterpart. While the catalogue tabulates
various multi-wavelength and derived properties, such as photometric redshifts, for the purposes of this work
we only use positions of optical/IR counterparts and integrated LoTSS radio flux measurements. 

We linked the SDSS QSOs to LoTSS sources using the counterpart positions from this catalogue.
All the QSOs in our working sample have {\it WISE} W1 detections (by construction; see Section \ref{wise_data}) and are bright enough
to be detectable in PanSTARRS. Therefore, we expect negligible incompleteness to be introduced into our sample
by using the associations from the LoTSS DR1 catalogue. Indeed, we found that essentially all cross-matches between the
positions of the radio source associations and those of our QSOs had offsets $< 0\farcs75$. Taking this as the
cross-matching threshold, we linked 1868 QSOs to LoTSS radio sources with an integrated S/N$>5$ 
($\approx 16$\% of the working QSO sample). In Section \ref{det_stats}, we present a more detailed analysis 
of the radio detection statistics of the QSOs.

\subsection{FIRST} \label{first_data}

We also searched for associations between QSOs from our working sample and
radio sources detected by the Faint Images of the Radio Sky at Twenty centimeters (FIRST) survey, a legacy
VLA programme that imaged the entire SDSS footprint at 1.4 GHz with a resolution of $5$'' 
and a 5$\sigma$ detection limit of 1 mJy \citep{becker95, helfand15}.
Following the approach taken by \citet{klindt19}, we assumed that all detected radio sources with centroids
within 10'' of a QSO are associated with it. In the cases of multiple radio sources within the 10'' search radius, 
we add the FIRST measurements of the integrated fluxes of all components to yield a total flux for such multi-component sources.
This simple criterion has a low false-match probability ($\approx 0.2$\%) but identifies 
the vast majority of extended and multi-component radio sources found among SDSS QSOs, missing only 
$\approx 2$\% of associations from large double-lobed radio sources \citep{lu07, klindt19}.

The LoTSS data offer a means to reliably recover the fluxes of very extended radio sources, owing to the high sensitivity
of the LOFAR images to low surface brightness steep-spectrum emission. The LoTSS DR1 associations catalogue
identifies radio sources that are composed of multiple components, and provides positions for each sub-component \citep{williams19}. Based
on our visual assessment of the LoTSS images (Section \ref{radio_morphs}), we determined that we could reliably associate
FIRST counterparts to these large extended sub-components using a secondary search tolerance of 3''. In this way, we obtained
more accurate total fluxes for 68 FIRST-detected radio QSOs, including emission well beyond 10'' from their central positions.

Of the QSOs in the working sample, only 544 ($\approx 4.6$\%) have a FIRST association. The difference in overall
detection rate with respect to LoTSS is a testament to the significantly better sensitivity of the LOFAR surveys 
to radio emission from QSOs. 

A small number of QSOs with FIRST detections do not have entries in the LoTSS DR1 association catalogue because
they have been excluded by the likelihood ratio matching technique due to complex structure or true mis-association.
We visually examined the 57 colour-selected QSOs from our working sample that fell into this category, and identified
reliable associations for 37. Only 20 QSOs are definite FIRST sources without a reliable LoTSS detection.
These QSOs are candidates for inverted spectrum radio sources (see Figure \ref{spectral_indices}), a rare and interesting population.

\subsection{Matching subsamples of colour-selected QSOs} \label{matching}

An important consequence of the differences in intrinsic luminosities between rQSOs and cQSOs (Section \ref{wise_data})
is the effects of {\it WISE} selection on the redshift distributions of the colour-selected QSO subsets. By construction, cQSOs and 
rQSOs in the parent sample have essentially identical redshift distributions: due to the non-uniformity of SDSS QSO selection across the sky,
there are subtle differences introduced by the restriction of the parent sample to the LoTSS survey area, but these are quite minor. 
Despite this, because the population of rQSOs are intrinsically brighter in the MIR, they become over-represented at 
higher redshifts after applying our {\it WISE} cuts. In the working sample, this effect leads to a long high redshift tail among 
rQSOs compared to cQSOs.

In order to minimise any biases in our results due to differences in redshifts and intrinsic luminosities of the two colour-selected
QSO subsets, we employ a procedure of matching the rQSOs to cQSOs of similar redshift and \lsix. For each rQSO
in the working sample, we randomly choose a unique cQSO within a redshift tolerance of $0.02\times(1+z)$ and 
a \lsix\ tolerance of 0.1 dex. 
We exclude from the analysis any rQSOs that do not have a matching cQSO from this process; 
in practice, however, $>99$\% of rQSOs are paired with cQSOs in any given iteration of the matching procedure. 
Henceforth, we refer to these matched subsamples as ``L6$z$-matched'' QSOs.

The matching procedure is a Monte-Carlo process. Due to its inherent randomness, all of the analysis in this work that relies on the L6$z$-matched QSO subsamples is subject to variance. We minimise the effects of variance in L6$z$-matching by performing a 
minimum of 30 independent iterations of each analysis. Results are quoted and figures are plotted using the median
statistics of these iterations. As an example, consider the lower panel of Figure \ref{radio_ztrends}. 
The blue points show radio detection
statistics of L6$z$-matched cQSOs in bins of redshift. The plotted blue stars are the median detection fractions from 30 iterations
of the matching procedure, and the errorbars on the plotted points show the median upper and lower errors using binomial
statistical calculations applied at each iteration, following the technique of \citet{cameron11}. A similar methodology
is used to portray summary statistics throughout the paper, unless specifically noted otherwise.


\begin{figure}
\centering 
\includegraphics[width=\columnwidth]{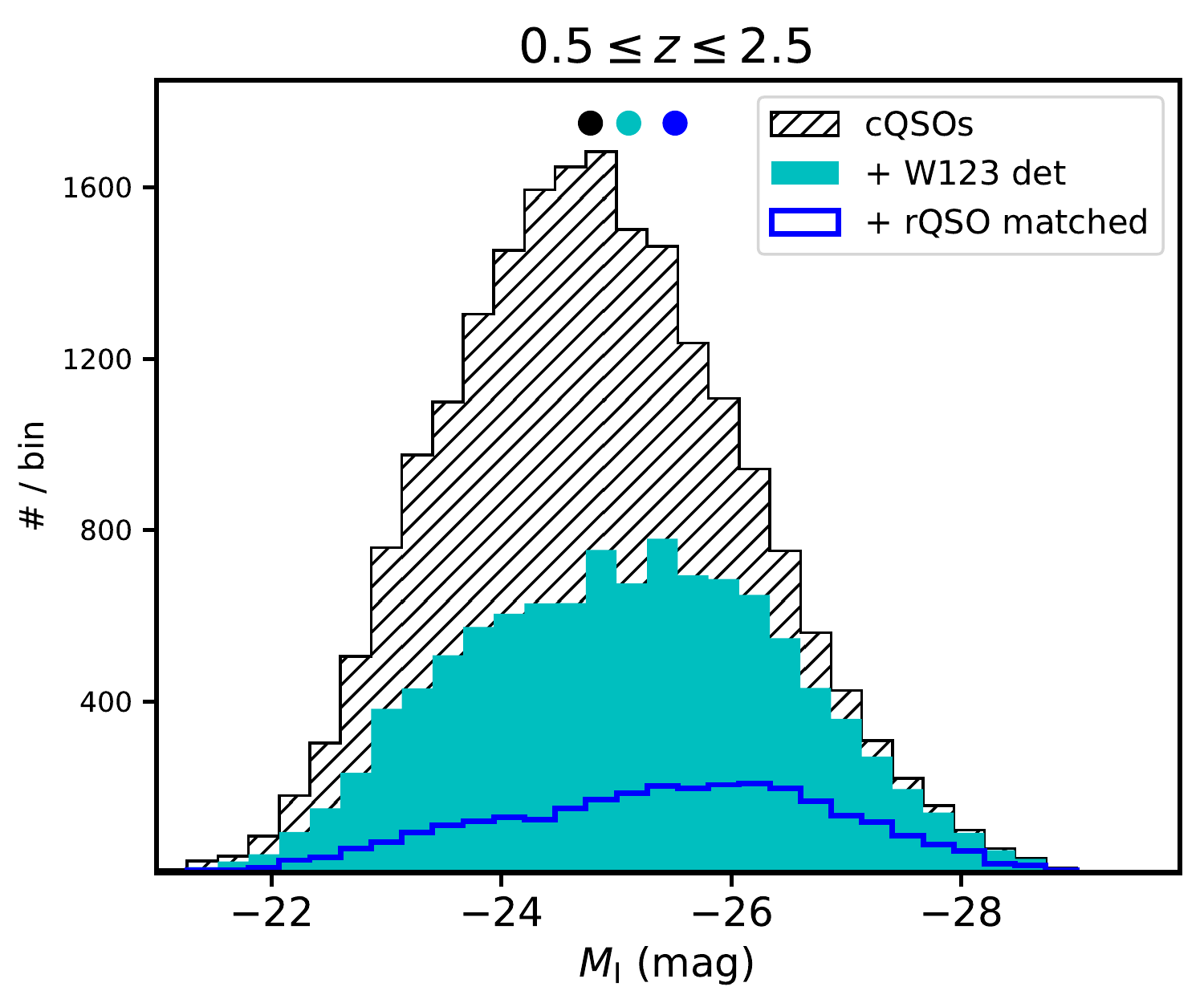}
\caption
{Distributions of rest-frame I-band absolute magnitude ($M_{\rm I}$) of cQSOs in the redshift interval
$0.5<z<2.5$. The hatched histogram shows all cQSOs from the SDSS DR14 parent sample, the filled cyan histogram
shows those cQSOs with the additional requirement of a SNR$>2$ measurement in the {\it WISE} 1, 2, and 3 bands,
and the open blue histogram indicates the cQSOs left after a further restriction imposed by matching to rQSOs in redshift and \lsix.
The black, cyan, and blue filled circular points at the top of the graph show the median $M_{\rm I}$ of the 
distributions in the respective order noted above.
cQSOs are typically less intrinsically luminous than rQSOs, and the L6$z$-matching process accounts for this.
We only show the results of a single iteration of L6$z$-matching in this Figure, but the blue histogram is essentially
unchanged by matching variance.
}
\label{fig2}
\end{figure}

In Figure \ref{fig2}, we illustrate how the steps in selection and matching alter the nature of the control QSOs used in this study.
The hatched black histogram shows the distribution of I-band absolute magnitudes, calculated by \citet{paris18},
of all cQSOs selected from the SDSS DR14 QSO
catalogue that lie in the region covered by LoTSS DR1 and in the redshift interval of $0.5<z<2.5$. With the requirement of a S/N$>2$
detection in the three shorter {\it WISE} bands (W1/2/3), the number of cQSOs drops by about 50\%, the remainder having a
median luminosity that is brighter by half a magnitude (cyan histogram). 
These constitute the cQSOs in the working sample. Finally, after matching to the
\lsix\ and redshifts of the rQSOs, we are left with an even smaller subset of cQSOs that are more luminous again by 0.3 mag (blue open histogram).

Figure \ref{fig2} demonstrates that as we settle towards a control set of normal QSOs that fully captures the joint distribution in redshift
and AGN luminosity of red QSOs, we select progressively more luminous systems. This highlights the importance of our fully
matched approach: we can be confident that any differences we see in the radio properties of rQSOs with respect to cQSOs
are not due to simple selection effects arising from luminosity or redshift mismatches.


\section{Results} \label{results}

\subsection{LoTSS detection statistics} \label{det_stats}

\begin{table*}
\caption{Radio detection statistics for SDSS DR14 colour-selected QSOs}
\label{det_stat_table}
\begin{tabular}{ccccc}
\hline 
\hline 
Subset & $0.5 \leq z < 0.8$ & $0.8 \leq z < 1.5$  & $1.5 \leq z < 2.0$ & $2.0 \leq z < 2.5$ \\
\hline 
\multicolumn{5}{c}{Working sample$^{\rm a}$: LoTSS DR1 detection statistics and detection fractions} \\ 
 \hline 
cQSOs &  158  / 1046  (14\% -- 16\%) & 505  / 3906  (12\% -- 13\%) & 334  / 2599  (12\% -- 13\%) & 195  / 1657  (11\% -- 12\%)\\
rQSOs &  67   / 285   (21\% -- 26\%) & 273  / 968   (26\% -- 29\%) & 177  / 687   (24\% -- 27\%) & 158  / 598   (24\% -- 28\%)\\\hline 
\multicolumn{5}{c}{L6$z$-matched sample$^{\rm b}$ (30 iterations): LoTSS DR1 detection fraction} \\ 
\hline 
cQSOs &  12\% -- 16\% & 14\% -- 16\% & 13\% -- 16\% & 12\% -- 15\%\\
rQSOs &  20\% -- 25\% & 26\% -- 29\% & 23\% -- 27\% & 24\% -- 27\%\\
\hline 
\multicolumn{5}{c}{L6$z$-matched sample$^{\rm b}$ (30 iterations): FIRST detection fractions} \\ 
\hline 
cQSOs &   2\% --  4\% &  3\% --  5\% &  2\% --  4\% &  2\% --  3\%\\
rQSOs &   4\% --  7\% &  7\% --  9\% &  6\% --  8\% &  7\% --  9\%\\
\hline 
\hline
\multicolumn{5}{l}{\footnotesize{Detection fractions, shown as percentages, span the 16th to 84th percentile confidence intervals.}}\\
\multicolumn{5}{l}{\textsuperscript{a}\footnotesize{Left / Right numbers are radio-detected / total}}\\
\multicolumn{5}{l}{\textsuperscript{b}\footnotesize{Absolute numbers are subject to L6$z$-matching variance. Therefore, only detection fractions are shown.}}\\
\end{tabular}
\end{table*}

The simplest comparison we can make between the 144 MHz radio properties of rQSOs and cQSOs
is in terms of their detection rates in LoTSS DR1. We consider a source to be detected if the integrated
S/N of the radio source (including all elements of a multi-component source) is greater than 5.

\begin{figure*}
\centering 
\includegraphics[width=\textwidth]{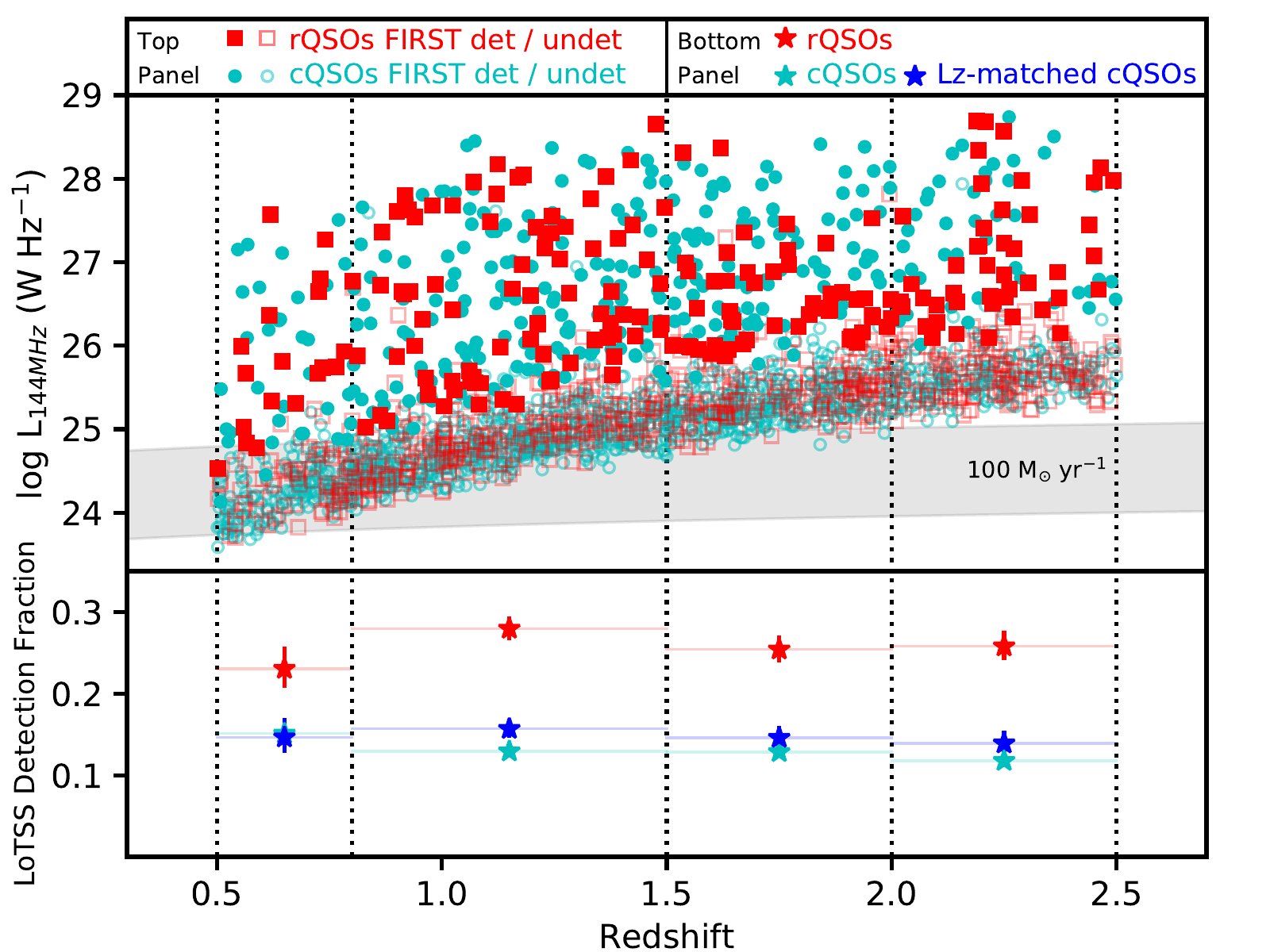}
\caption
{Upper panel: The meter-wave radio luminosity (\lrad) against redshift for QSOs from the working sample. Red/cyan
coloured points represent rQSOs/cQSOs. Filled points are those sources detected with S/N$\geq 5$ in both LoTSS DR1
and FIRST surveys. The grey band shows the range in \lrad\ from star-forming galaxies with a SFR of 100 \msun\ yr$^{-1}$
based on the FIR-radio correlation and its evolution from \citet{calistro17}. Lower panel: The LoTSS DR1 detection
fractions for rQSOs (red points), cQSOs from the working sample (cyan points), and cQSOs from the L6$z$-matched sample
(blue points). The fractions are evaluated for QSOs in four redshift bins, delineated by vertical dotted lines in both panels.
}  
\label{radio_ztrends}
\end{figure*}

 In Table \ref{det_stat_table}, we show
the fraction of rQSOs and cQSOs in each of our four primary redshift bins that are detected in LoTSS DR1. We report
separate detection fractions for the working sample and the L6$z$-matched subsample. Errors
in the detection fractions are calculated using the methodology of \citet{cameron11}. 
These detection fractions are also plotted in the lower panel of Figure \ref{radio_ztrends}, where
we only show the points for L6$z$-matched rQSOs since they are practically the same objects
as the rQSOs from the working sample.

We find that rQSOs are detected at a substantially higher rate in LoTSS compared to the cQSOs.
The detection fraction of rQSOs remains constant at $\approx 25$\% across all redshifts considered in this study,
and is always higher than the $\approx 13$\% detection fraction of the cQSOs. The effect of matching in \lsix\ and redshift
only marginally increases the cQSO detection fraction, confirming the conclusion from \citet{klindt19}
that differences in intrinsic accretion rate are not the primary determinant of this result.

Since the FIRST 1.4 GHz survey covers the LoTSS DR1 area, we can also compare the FIRST detection statistics of the
colour-selected QSOs between the two radio datasets. Table \ref{det_stat_table} also lists the
fractions of L6$z$-matched colour-selected QSOs that have S/N$\geq 5$ detections in the FIRST survey. At all redshifts, these
fractions are consistently factors of $\sim3$ lower than the detection fractions in LoTSS DR1 for the same sample of QSOs.

LoTSS is considerably more sensitive than FIRST to the emission from radio QSOs across a wide range of spectral
indices. Only FIRST sources with strongly inverted metre-wave spectra, or those that display strong variability on decade
time-scales, will have no LoTSS DR1 counterpart in the overlapping area of the two surveys. After visually reconciling
mis-associated sources (see Section \ref{first_data}), and excluding FIRST sources at the edges of the LoTSS maps, we find 
that $\approx 97$\% of FIRST-detected QSOs are also detected in LoTSS. In Section \ref{lradio_dists}, we take advantage of the mutual 
completeness of LoTSS and FIRST to examine and compare the radio spectral indices of colour-selected QSOs.

LoTSS offers us an important opportunity to evaluate the radio selection biases of SDSS rQSOs. 
Since FIRST predated the bulk of the SDSS Legacy spectroscopic survey, radio detections were used as a way to distinguish 
between stars and QSOs in the early targeting strategy \citep{richards02}. This pre-selection of radio-bright QSOs, particularly
in parts of the QSO multi-colour selection space with low completeness (overlapping the colours of stars or at high redshifts),
will preferentially boost the incidence of radio QSOs with abnormal colours, 
such as rQSOs. \citet{klindt19} minimise such biases by only studying
SDSS DR7 QSOs that satisfy the principal multi-colour selection of the Legacy SDSS survey, i.e., those flagged as \texttt{UNIFORM} 
in \citet{schneider10}. The SDSS DR14 QSO catalogue is much larger and much more heterogeneous than the DR7 edition.
rQSOs selected from this catalogue are therefore more susceptible to FIRST pre-selection biases.

However, the LoTSS survey pushes into a regime of radio luminosity that was not covered by FIRST (see Section \ref{lradio_dists}).
As a much more recent survey, the sources in LoTSS are not entangled with the selection function of SDSS DR14 QSOs.
Even if we exclude all FIRST-matched QSOs from the L6$z$-matched sample, effectively removing any residual bias
introduced by FIRST pre-selection, we still find that rQSOs have $\times1.6$ higher LoTSS detection rates compared to cQSOs.
The higher proportion of rQSOs in LoTSS are an important and independent validation of the key result from \citet{klindt19}: 
the population of red QSOs do produce significantly larger amounts of detectable radio emission than normal QSOs.

\subsection{Radio luminosities} \label{lradio_dists}

\begin{figure}
\centering 
\includegraphics[width=\columnwidth]{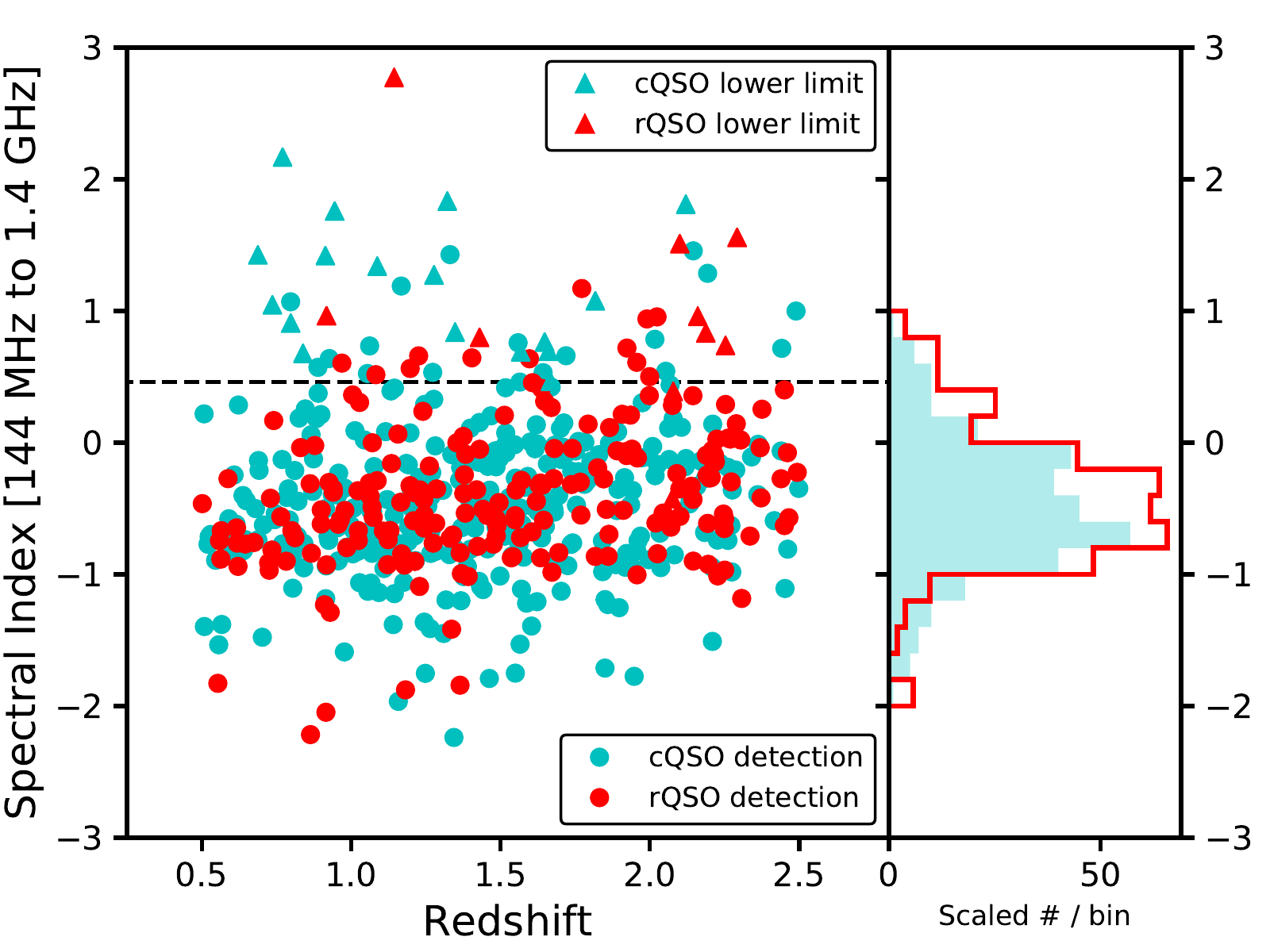}
\caption
{Radio spectral indices between 144 MHz and 1.4 GHz ($\alpha_{\rm R}$, defined such that 
the monochromatic radio luminosity $L_{\nu} \propto \nu^{\alpha_{R}}$) 
for all FIRST-detected QSOs from our working sample.
Red/cyan colours represent rQSOs/cQSOs. Left panel: $\alpha_{\rm R}$ vs. redshift. 
The dashed line is the spectral index of a source that lies at the nominal $5\sigma$ detection limits of both FIRST and LoTSS DR1.
Only FIRST-detected sources with very flat or inverted radio spectra can be too faint to be detected by LoTSS.
Right panel: Distributions of $\alpha_{\rm R}$ for QSOs detected in both surveys.}  
\label{spectral_indices}
\end{figure}

A comparison of the distributions of 144 MHz radio luminosities (\lrad) between the two colour-selected QSO subsets can inform
us on the possible causes for the higher incidence of radio emission among rQSOs. 
Because of the large range in redshifts covered by our working sample, the calculation of monochromatic radio
luminosities requires the application of a k-correction to the observed-frame LoTSS fluxes to bring them to a
fixed rest-frame estimate. We take the standard approach of assuming a single spectral slope for the meter-wave
radio continuum emission parameterised by the spectral index $\alpha_{R}$, such that
the monochromatic radio luminosity $L_{\nu} \propto \nu^{\alpha_{R}}$.

For the QSOs with detections in both LoTSS DR1 and FIRST, we calculate a representative $\alpha_{R}$
from the ratio of the integrated fluxes at 144 MHz and 1.4 GHz. In Figure \ref{spectral_indices},
we plot this measured spectral index against redshift, differentiating between rQSOs and cQSOs. The spectral
indices show a broad distribution with a peak at $\alpha_{\rm R} \approx -0.7$, a typical value for the 
cosmic synchrotron-dominated radio source population.

The vast majority of LoTSS detected QSOs are too faint to be detected in FIRST. For these objects, we
estimate \lrad\ assuming a fixed $\alpha_{R} = -0.7$. We have verified that none of the results presented
in this paper depend on the exact value for the spectral index assumed for the FIRST-undetected radio sources.

In the upper panel of Figure \ref{radio_ztrends}, we plot \lrad\ against redshift for all LoTSS detected QSOs
from the working sample, split by colour-selection. Filled/open symbols indicate which sources are detected/undetected
in the FIRST survey. Interestingly, there is a substantial increase in the number counts of radio sources just below the
FIRST limit, indicating that LoTSS is probing a different regime of the radio QSO population that was not well-characterised
by FIRST \citep[e.g.,][]{gurkan19}. A similar result has been noted at higher frequencies from deep targeted
radio observations of QSOs \citep[e.g.,][]{kimball11}, and can be understood as an increasing proportion of 
star-formation powered systems among low luminosity radio QSOs.
We will explore the nature of this population in more detail in Section \ref{sf_origins}.

\subsection{Radio loudness} \label{R_results}

\begin{figure}
\centering 
\includegraphics[width=\columnwidth]{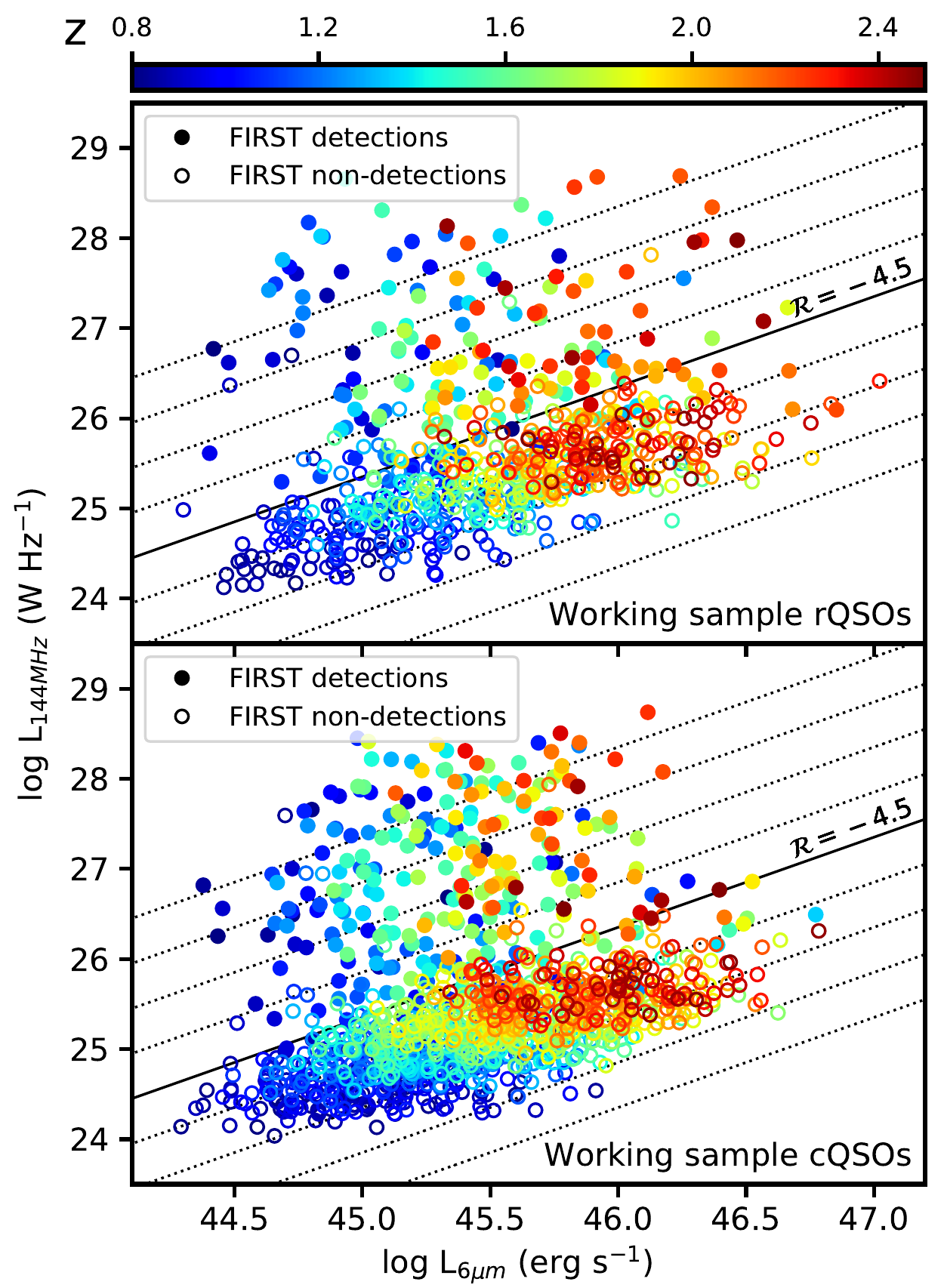}
\caption
{Metre-wave radio luminosity (\lrad) vs. 6 \mics\ MIR luminosity (\lsix) for cQSOs (left) and rQSOs (right).
The colours of the plotted points indicate their redshifts (see colourbar). 
FIRST-detected sources are shown with filled symbols, while open symbols show sources that are only
detected in LoTSS. The black solid line marks a constant `radio-loudness' of $\mathcal{R} = -4.5$, the approximate
threshold between radio-loud and radio-quiet systems. Dotted black lines show different values of 
$\mathcal{R}$ that are 0.5 dex apart, and the system of black lines delineate bins of $\mathcal{R}$ featured
in Figure \ref{detratioR}.
}  
\label{mir_radio}
\end{figure}

Additional insight into the nature of this population comes from the relationship between MIR and radio power,
which we examine in Figure \ref{mir_radio} by plotting \lrad\ against \lsix\ for LoTSS-detected 
QSOs at $0.8<z<2.5$. We exclude the lowest redshifts, those from the working sample at $0.5<z<0.8$,
to minimise any contamination from strong starbursts (see Section \ref{wise_data} and Figure \ref{l6comp}).
We treat cQSOs and rQSOs separately, and distinguish between those with FIRST
detections (filled symbols) and those without (open symbols). 

Both cQSOs and rQSOs span a wide range in \lsix\ and \lrad. The FIRST-detected systems, being brighter in the radio, occupy
the upper scatter of radio luminosities (filled symbols), while those detected solely in LoTSS are inherently less radio-luminous. 
Among these weaker sources (open symbols), a correlation appears to exist between \lsix\ and \lrad. However, based on
an examination of their redshifts (the colour scheme used to plot the points in  Figure \ref{mir_radio}), 
we conclude that the correlation is driven primarily by the strong redshift-dependent luminosity 
bias of our sample. At any given redshift, the trend between the MIR and radio luminosities is not 
obvious, suggesting at best a weak intrinsic correlation.

The diagonal lines in Figure \ref{mir_radio} delineate bins of constant \lrad--\lsix\ ratio. As we can be assured that \lsix\ for
SDSS QSOs at these redshifts is dominated by reprocessed AGN radiation (Section \ref{wise_data}), this ratio is a measure of the relative power
of synchrotron emission to accretion disc emission, often referred to as `radio-loudness' ($\mathcal{R}$) in the common literature.
For our purposes, we define $\mathcal{R}$ in dimensionless fashion as follows:

\begin{equation}
\mathcal{R} \; = \; \log \; \frac{1.44\times10^{13} \, L_{\rm 144}}{L_{\rm 6 \mu m}} 
\label{eqn1}
\end{equation}

\noindent with \lrad\ in W Hz$^{-1}$ and \lsix\ in \ergs.

Radio-loudness has a number of different definitions in the literature. Traditionally, it is expressed as a ratio of rest-frame 
luminosity densities \citep[e.g.][]{kellerman89} or even observed-frame flux densities \citep[e.g.][]{ivezic02}. For such
definitions, the typical divide between `radio-loud' and `radio-quiet' systems has a value of $\approx 10$. For our
purposes, we instead use a ratio of rest-frame monochromatic powers (or luminosities per logarithmic frequency interval),
which is a better representation of the relative amount of energy released in the synchrotron component by the AGN (see Section 4.2).

In \citet{klindt19}, a boundary value was derived between radio-loud and radio-quiet systems for the ratio of 1.4 GHz and 
{\it WISE}-based MIR luminosities, calibrated to be equivalent to the more traditional divide laid out by early SDSS studies \citep{ivezic02}. 
As our study deals with 144 MHz luminosities, we translate the calibration from \citet{klindt19} by assuming a radio spectral index of -0.7, 
which results in a boundary at a value of $\mathcal{R} = -4.5$. A source with a canonical spectral index that lies on this boundary 
in our study will also lie on the equivalent boundary at 1.4 GHz, but would have radio-loudness parameters of $-4.2$ in the works of
\citet{klindt19} and \citet{fawcett20}.

\begin{figure*}
\centering 
\includegraphics[width=\textwidth]{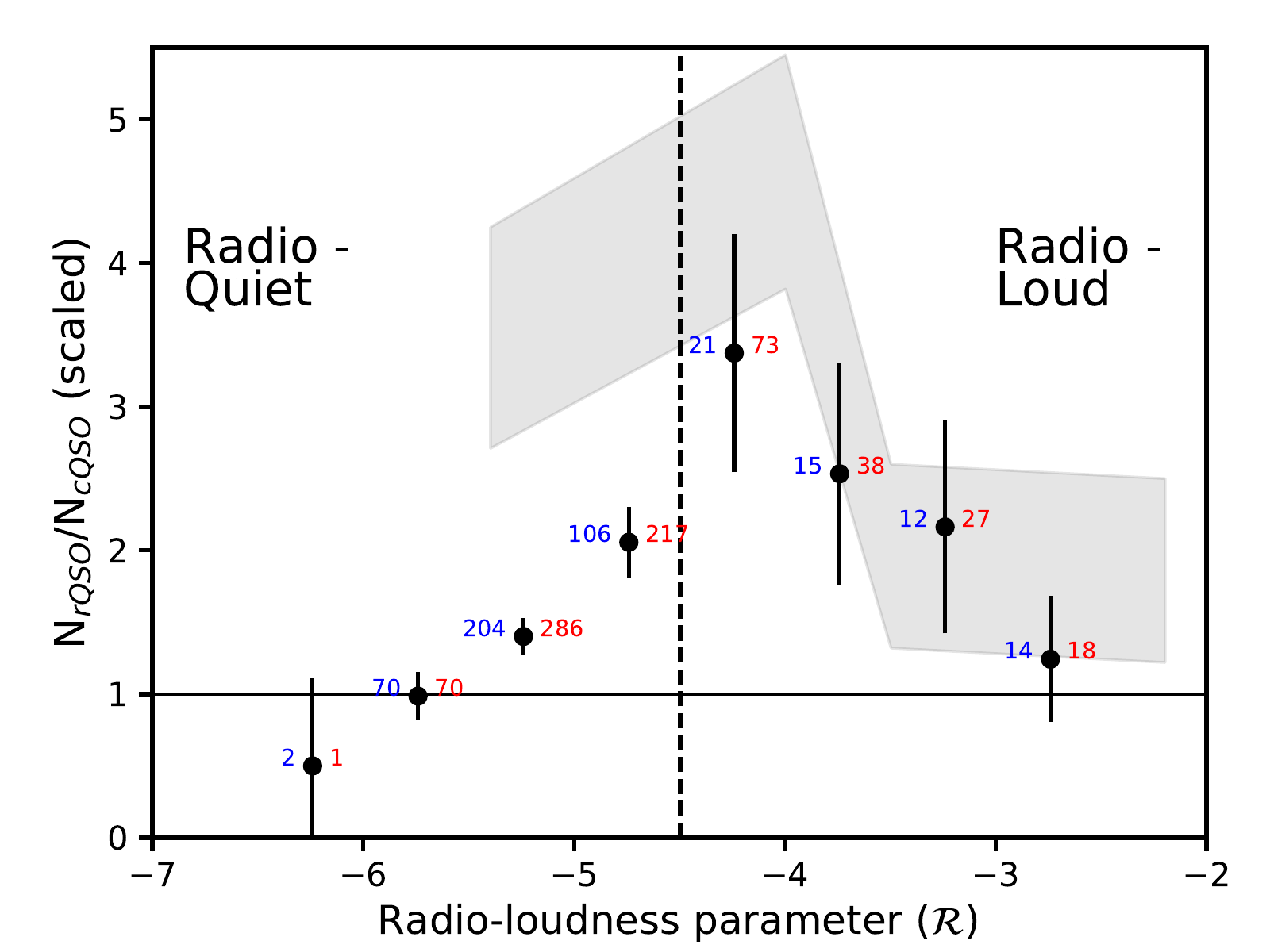}
\caption
{The ratio of the detection rate in LoTSS for rQSOs vs. L6$z$-matched cQSOs plotted in bins of the `radio-loudness'
parameter $\mathcal{R}$, defined in Equation \ref{eqn1} as the dimensionless ratio of the MIR 
and meter-wave radio luminosities. The coloured numbers next to the points indicate the number of 
LoTSS-detected rQSOs (red) and cQSOs (blue) that contribute to the ratio, a median from 30 bootstrap draws of
L6$z$-matching.
The vertical dashed line at $\mathcal{R} = -4.5$ marks the threshold between systems that are
defined as `radio-loud' (to the right) and `radio-quiet' (to the left). Note the broad peak around this divide at which
rQSOs are more common than cQSOs among LoTSS radio sources.
The grey filled polygon shows the detection rate ratio for rQSOs vs. L6$z$-matched cQSOs 
from the FIRST-based study of \citet{klindt19}, appropriately shifted by $\Delta \mathcal{R}$ = -0.3 dex to 
account for the differences between the 1.4 GHz and 144 MHz bands.
}  
\label{detratioR}
\end{figure*}

In Figure \ref{mir_radio}, we plot lines of constant $\mathcal{R}$ as diagonal lines, incremented
in steps of 0.5 dex. The solid black line in Figure \ref{mir_radio} marks $\mathcal{R} = -4.5$ 
the separation between radio-loud and radio-quiet systems discussed above.
Taking this fiducial separation, we can compare the relative number of radio-loud and radio-quiet
QSOs in both colour-selected subsets. We find that 24\% of LoTSS-detected rQSOs are radio-loud, as opposed
to $\approx 18$\% of L6$z$-matched cQSOs, a small but significant difference.

Following \citet{klindt19}, we take this analysis further by examining whether the relative enhancement of the radio detection 
rates of rQSOs over cQSOs depends on $\mathcal{R}$. In Figure \ref{detratioR}, we plot the ratio of LoTSS-detected
rQSOs and L6$z$-matched cQSOs in the same bins of $\mathcal{R}$ as set forth in Figure \ref{mir_radio}. 
At any value of $\mathcal{R}$, a value of this ratio $>1$ indicates that rQSOs with 
that degree of `radio-loudness' are preferentially detected in LoTSS. 
We find a definite peak in the Figure at $\mathcal{R} \approx -4.5$ spanning the divide between radio-quiet and radio-loud sources. At
both large and small values of $\mathcal{R}$, rQSOs are as likely to be detected in LoTSS as cQSOs, but they are $\approx 3$ times
more numerous than cQSOs if they are of intermediate radio-loudness. 

The grey filled polygon in Figure \ref{detratioR} shows the detection rate enhancement among L6$z$-matched rQSOs and cQSOs
from the FIRST-based of \citet{klindt19}. Their data have been shifted along the $\mathcal{R}$ axis by 0.3 dex to account for the
differences in the radio bands between FIRST and LoTSS, allowing a comparison to our own data points. The rise
in the detection rate enhancement with decreasing $\mathcal{R}$ is completely consistent between our two studies,
but we find a sharper and more significant dropoff among radio-quiet systems. Given the differences in the QSO parent
samples, the redshift range under consideration, the statistical methodology,
and the fact that FIRST is only barely sensitive enough to detect radio-quiet AGN
(Figure \ref{mir_radio}), we consider the comparison between our two studies to be quite favourable.

Due to the much-improved sensitivity afforded by LoTSS over FIRST, our analysis is able to track the transition 
clearly for the first time and identify that the enhanced radio emission in rQSOs is preferentially found among radio-intermediate
QSOs. In the parallel work of \citet{fawcett20}, a similar result is evident from a joint analysis of the Stripe82 and COSMOS radio surveys.

\subsection{Radio source morphologies} \label{radio_morphs}

\begin{figure*}
\centering 
\includegraphics[width=\textwidth]{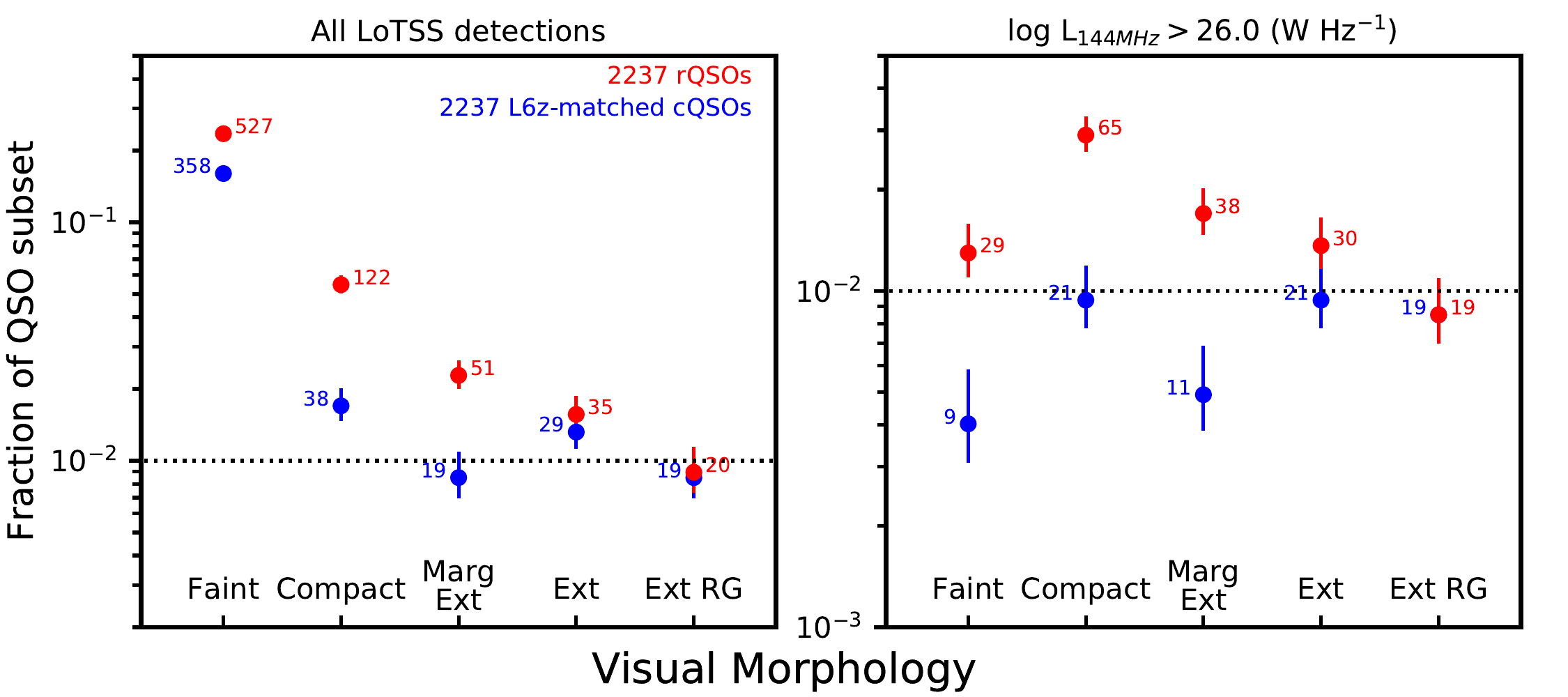}
\caption
{Radio morphology distributions of L6$z$-matched rQSOs (red) and cQSOs (blue).
The Y-axis is the fraction of QSOs of a certain morphology (marked on the X-axis) out of the total number of
QSOs of the colour-selected subset in the redshift range $0.8<z<2.5$. 
30 bootstrap draws of L6$z$-matching were used to generate the data for this Figure.
The coloured numbers next to the points indicate the median number of radio QSOs that contribute to the fraction, 
while the coloured numbers on the top right of each panel give the median of the total number of QSOs
analysed from all the draws. See Section \ref{radio_morphs} for a description of the morphological
classes. The left panel considers the distributions for all LoTSS detections, while the right considers only
the subset with a meter-wave radio luminosity \lrad $ > 10^{26}$ W Hz$^{-1}$ (see Section \ref{synthesis}
for the associated discussion).
}  
\label{vismorphs}
\end{figure*}

The observed morphologies of radio QSOs are determined by  
the power and accretion mechanisms of the central engines, their orientation to the line-of-sight,
as well as their environments \citep{miley80,best12}. 
Through an examination of the range of 144 MHz morphologies exhibited by the colour-selected
QSOs, we expect to gain some insight into the nature of the excess radio emission found in rQSOs.

QSOs with an integrated LoTSS S/N $= 5$--$15$, as determined
from integrated flux measurements in the LoTSS DR1 associations catalogue \citep{williams19},
are often too faint for a reliable visual classification. Some lie in regions of the LoTSS field 
which have not been deconvolved sufficiently for a good image. 
In our morphological analysis, we place these sources into a `Faint' category determined purely by their overall S/N.

We refer to the remaining LoTSS radio sources from the working sample with S/N $>15$ as the `Bright' subset,
and these were the targets of our assessment of visual morphology. By their selection, they comprise a more
luminous subset of the colour-selected radio QSOs. However, we have verified that the LoTSS detection 
fraction of rQSOs in the Bright subset (11\%) remains high compared to cQSOs in this subset ($\approx 5$\%), and
we can therefore consider them representative of the full working subsample.

\begin{figure*}
\centering 
\includegraphics[width=\textwidth]{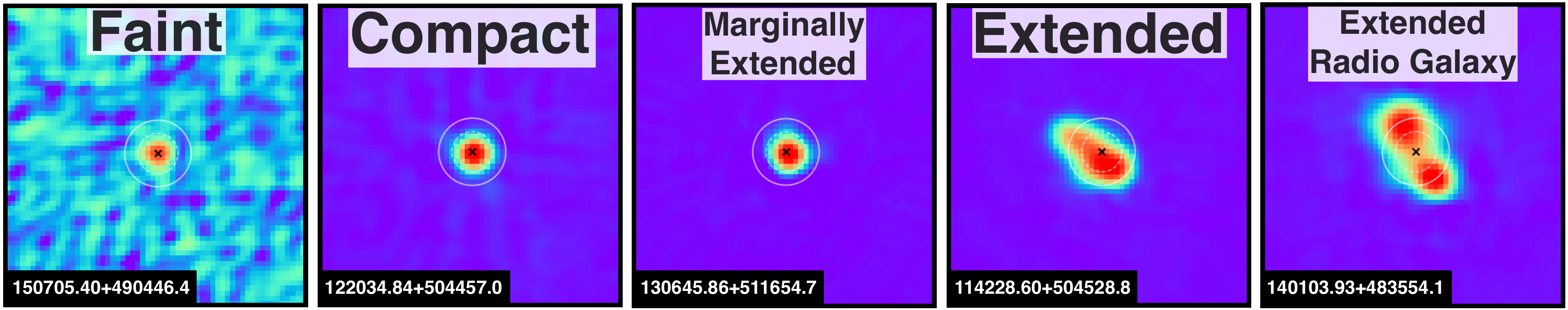}
\caption
{Typical examples of the radio morphology classifications that are used in the morphological analysis of Section \ref{radio_morphs}.
The QSOs featured here all lie at $1.0<z<1.5$.
Each image is 1.5 arcminutes on a side and oriented North up and East to the left. The visual category is written in large black letters
at the top of each panel, and the SDSS ID of the respective QSO, which incorporates its celestial position, is written 
in white letters at the bottom left. The concentric white circles provide reference diameters of 6'' and 10'' centred on the QSO position.
The `Faint' category is not a morphological category itself, but includes all LoTSS sources that have
a S/N $< 15$ from the survey catalogue. 
}  
\label{morph_examples}
\end{figure*}

For the purposes of visual morphology classification, we created 
1.5 arcminute square cutouts centred on the SDSS positions of the 652 QSOs from the Bright subset, 
and visualised them using a large dynamic range stretch. We assessed the cutout images guided by
reference circles with diameters of 6'' (the generalised PSF width of LoTSS) and 20'' (within which sources are considered 
single associations in FIRST). Figure \ref{morph_examples} features the stretch, colour table, and overall layout of
the images used for classification. 

The LoTSS sensitivity varies across the field and around bright sources, which could translate into
correlations in the appearance of the noise level in the images if they were examined in the
order of a catalogue sorted by sky coordinates. To avoid this, we randomised the order of the stamps 
before visual classification.

We classified the radio source in each stamp into four categories:

\begin{itemize}

\item {\bf Compact:} The source size is visually consistent with the LoTSS average PSF and there are no other nearby
clearly associated lobes.

\item {\bf Marginally Extended:} The source is mildly extended beyond the LoTSS beam, with all of its flux contained within a 10''
radius of the QSO's position. The source's extended appearance or asymmetry may be due to factors such as local
noise, the non-uniformity of the local PSF in different parts of the LoTSS DR1 area, or due to genuine source extension
at the scale of 6''. We treat this category as a transition between compact and well-extended morphologies; it likely
contains a mixture of radio sources of both types.

\item {\bf Extended:} The source is clearly extended at the resolution of the LoTSS images, with emission apparent beyond
10'' of the QSO's position. However, it does not demonstrate the classical bi-polar geometry typical of radio galaxies. 
This category includes sources with a diverse range of sizes and structures, spanning the divide between compact
radio morphologies and some of the largest radio sources in LoTSS. It may also include highly asymmetrical
radio galaxies in which only one of the jets or lobes is bright enough to be visible in the LoTSS images.

\item {\bf Extended Radio Galaxy:} The source clearly shows a bi-polar jet/lobe geometry with the QSO at the radio core. These are often
the largest and brightest radio QSOs.

\end{itemize}

Figure \ref{morph_examples} show typical examples of the morphology categories, including an example of `Faint' class for which
no explicit classification was made. 

In the left panel of Figure \ref{vismorphs}, we plot the fraction of colour-selected QSOs in each visual category relative to all QSOs
in the respective colour-selected subset. Only L6$z$-matched QSOs in the redshift range of $0.8<z<2.5$ are considered here so
we can be confident that AGN luminosity-driven selection effects are minimal.

We find a significant difference in the incidence of faint, compact and marginally extended radio sources among rQSOs
compared to equally luminous cQSOs. About $1-2$\% of cQSOs are found in each of the four visually-assessed categories. 
In contrast, $\approx 6$\% of rQSOs show compact radio morphologies, a factor of 3 times higher than cQSOs. In essence,
almost all of the difference in the LoTSS detection rates of rQSOs arises through an excess of compact or faint radio sources. 

This result independently confirms the findings of \citet{klindt19}, in which a similar morphological analysis was performed
on FIRST images of SDSS DR7 QSOs. With the deeper images and better sensitivity to extended emission offered by LoTSS DR1, 
and the inclusion of fainter QSOs from the SDSS DR14 catalogue, we probe a different regime of accretion and radio power among
QSOs. The similarity of the results from both works is decisive validation of the special radio properties of red QSOs.

\subsection{Radio source properties as a function of morphology} \label{radio_alphas}

\begin{figure*}
\centering 
\includegraphics[width=\textwidth]{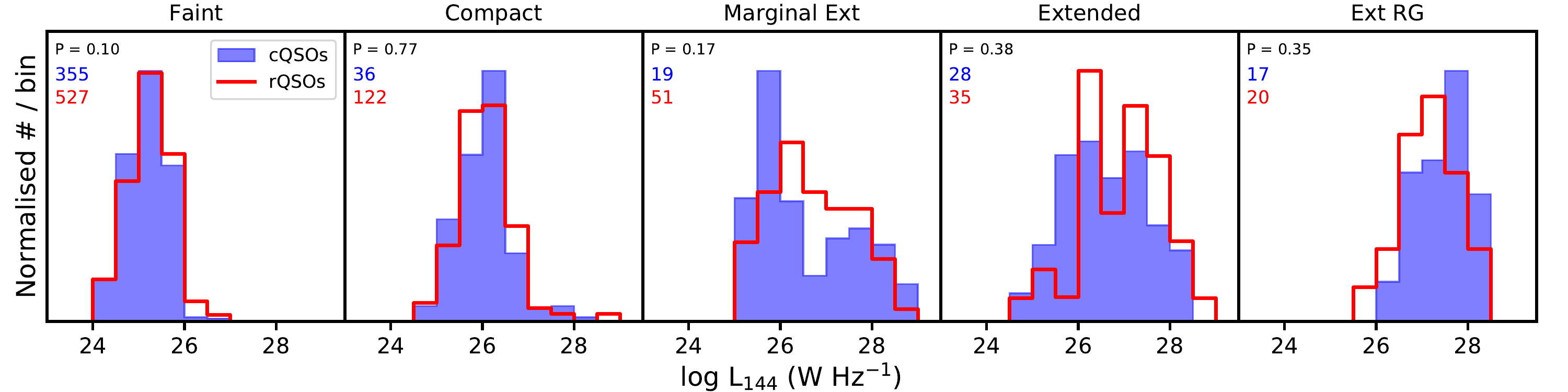}
\caption
{Comparisons of the normalised distributions of metre-wave radio luminosity (\lrad) of L6$z$-matched rQSOs (red open histograms)
and cQSOs (blue filled histograms), separated into the five morphological categories (see Section \ref{radio_morphs}).
Only LoTSS-detected QSOs at $0.8<z<2.5$ are considered. 
30 bootstrap draws of L6$z$-matching were used to generate the data for this Figure.
The numbers on the top left of each panel are the median number of colour-selected QSOs that comprise
the histograms. The median K-S {\it P} value is written in the top-left corner of each panel:
this is the probability that the distributions are drawn from the same parent population, a measure of their similarity. 
}  
\label{morph_lrdists}
\end{figure*}

\begin{figure*}
\centering 
\includegraphics[width=\textwidth]{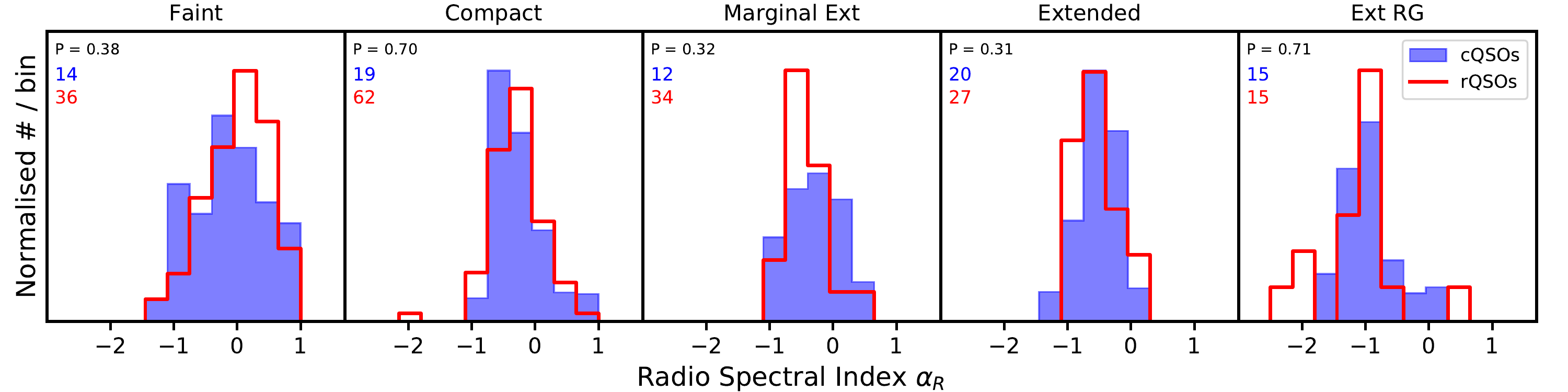}
\caption
{Comparisons of the normalised distributions of radio spectral index ($\alpha_{R}$) of L6$z$-matched rQSOs (red open histograms)
and cQSOs (blue filled histograms), separated into the five morphological categories (see Section \ref{radio_morphs}).
Only LoTSS-detected QSOs at $0.8<z<2.5$ that are additionally detected in the FIRST survey are considered here. 
30 bootstrap draws of L6$z$-matching were used to generate the data for this Figure.
The numbers on the top left of each panel are the median number of colour-selected QSOs that comprise
the histograms. The median K-S {\it P} value is written in the top-left corner of each panel:
this is the probability that the distributions are drawn from the same parent population, a measure of their similarity.}  
\label{morph_alphadists}
\end{figure*}

With a complete set of morphological classifications for the Bright colour-selected QSOs, we are able to examine
the properties of their radio sources as a function of morphology. We place particular emphasis on the 
`Compact' category since it is among these sources that rQSOs differ most from cQSOs.

In Figure \ref{morph_lrdists}, we compare the distributions of radio luminosity for both subsets of colour-selected QSOs in the L6$z$-matched
sample, split by morphology and including Faint sources as well. In this Figure, and also for Figure \ref{morph_alphadists} described below, 
we aggregate QSOs over the entire $0.8<z<2.5$ redshift interval to improve overall number statistics, rather than divide them
into redshift bins.
Therefore, the distributions plotted are subject to redshift-dependent systematics. However, because of the use of redshift-matching, 
we expect such systematics to apply equally to both cQSOs and rQSOs, allowing a fair comparison to be made between them.

For both colour-selected subsets, 
we find that the \lrad\ distributions change substantially from compact to extended sources (from left to right in Figure \ref{morph_lrdists}).
As expected, the Extended Radio Galaxy category encompasses the most luminous radio sources in our sample.
Despite this, both subsets of QSOs appear to show very similar \lrad\ distributions within each morphology category. 
We have verified this quantitatively using two-sided K-S tests, assuming that a significant difference would yield
a K-S {\it P} value $< 0.05$.

For the subset of QSOs with 1.4 GHz detections from FIRST, we can also examine the 
radio spectral indices ($\alpha_{R}$) as a function of
morphology, as shown in Figure \ref{morph_alphadists}.
We find that the $\alpha_{R}$ distributions between the two subsets of 
colour-selected QSOs are indistinguishable (K-S {\it P} values $> 0.05$), 
especially among compact radio sources where we have the best statistics.
We conclude, therefore,that rQSOs with compact radio sources, while 
significantly more common than normal cQSOs, are not uniquely different in terms
of either their intrinsic radio luminosities or their radio spectra.

\section{Discussion} \label{discussion}

We have combined the SDSS DR14 catalogue of QSOs and the LoTSS DR1 meter-wave radio survey to undertake a
controlled study of red QSOs to $z=2.5$. Our strategy is to compare the radio properties of rQSOs (the 10\%
of QSOs with the reddest $g-i$ colours at any redshift) with typical cQSOs (the 50\% of QSOs around the median
$g-i$ colour at any redshift), matched in both redshift and mid-infrared luminosity (Section \ref{matching}).

LoTSS DR1 detects radio sources that are almost an order of magnitude less luminous than the FIRST survey, allowing
us to explore the properties of a different population of red QSOs than earlier work from our team \citep{klindt19}. The depth of LoTSS
also allows us to overcome the effects, however minor, of biases introduced into our sample by the FIRST-preselection of QSOs
in SDSS.

Our study confirms the results of \citet{klindt19}, many of which are also independently verified in the recent
work of \citet{fawcett20}.
At all redshifts considered in our study, the population of rQSOs are associated with an enhanced level of
detectable radio emission compared to cQSOs, by a factor of $\approx 2$ (Section \ref{det_stats}).
The enhancement is found among a preferred subset of rQSOs, those that are radio-quiet, with a radio-to-MIR
luminosity ratio ($\mathcal{R}$) in the range of $-5$ to $-3$ (Section \ref{R_results}), 
and display radio morphologies that are compact at the LoTSS spatial resolution (Section \ref{radio_morphs}).
Despite their preponderance among rQSOs, these compact, radio-quiet radio sources do not distinguish
themselves in terms of their radio spectral index over a decade in frequency (between 1.4 GHz and 144 MHz; 
Section \ref{radio_alphas}).

In the remainder of our discussion, we explore the origin of the meter-wave radio emission in SDSS QSOs.
We start with a set of Monte-Carlo simulations that use the well-known 
``FIR-radio correlation'' to constrain the synchrotron emission powered by 
star formation (Section \ref{sf_origins}). 
We then compare the observed radio emission to that expected given the SMBH masses and accretion rates
of the AGN in the QSOs. This will set the stage for a deeper understanding of the
possible origins of the particular radio properties of rQSOs, leading to our conclusive statements.

\subsection{The origin of meter-wave radio emission in QSOs: star formation} \label{sf_origins}

In the merger-driven evolutionary paradigm, red QSOs represent a phase of SMBH growth that parallels
the final stage of a massive gas-rich major merger and is co-eval with, or immediately follows, a massive starburst. 
In this context, one may expect rQSOs to be commonly found in hosts with very high levels of SF.
Since this SF is also capable of producing low-frequency radio emission \citep[e.g.,][]{calistro17, gurkan18, wang19},
it is valuable to test whether the higher incidence of compact radio sources among rQSOs may be produced by an underlying 
population of extreme starbursts that boost their radio power. 

To set the stage, we consider empirical measurements of the 144 MHz emission from SF
from the LOFAR study of \citet{calistro17}.
The cosmic rays accelerated in supernova explosions are 
responsible for the synchrotron emission from star-forming galaxies, and
this component correlates well with the SF-heated IR luminosity \citep[the ``FIR-radio correlation'', e.g.,][]{condon92}. 
From a set of star-forming galaxies with extensive multi-wavelength
data in the Bo\"otes field, \citet{calistro17} estimated the SFR using spectral energy distribution fits, and
calibrated the scatter and redshift evolution of the relationship between SFR
and \lrad. As a visualisation of this relationship, we plot a shaded band in 
Figure \ref{radio_ztrends} showing the expected range in \lrad\ for star-forming galaxies
with SFR $= 100$ \msun\ yr$^{-1}$, equivalent to local ultraluminous infrared galaxies (ULIRGs). The \lrad\ range
evolves with redshift since the nominal zeropoint of the FIR-radio correlation has a redshift dependence \citep{calistro17}.

Based on Figure \ref{radio_ztrends}, we note that a substantial subset of LoTSS-detected QSOs at $z\lesssim1$
could be powered by extreme starbursts (SFRs $> 100$ \msun\ yr$^{-1}$). Among the more distant
QSOs, SFRs would have to be an order of magnitude higher to substantially contribute to the emission of most of the
radio sources. Such systems would be among the most luminous star-forming galaxies in the Universe, 
the extremely rare hyperluminous infrared galaxies. Nevertheless, the 400 deg$^2$ of sky covered by LoTSS DR1 
samples a very large cosmic volume, so the field could cover a sufficient number of such rare systems to potentially 
account for the nature of the observed QSO population. 

%

\subsubsection{Simulating the SFR contribution in typical QSOs} \label{sfsims}

\begin{figure}
\centering 
\includegraphics[width=\columnwidth]{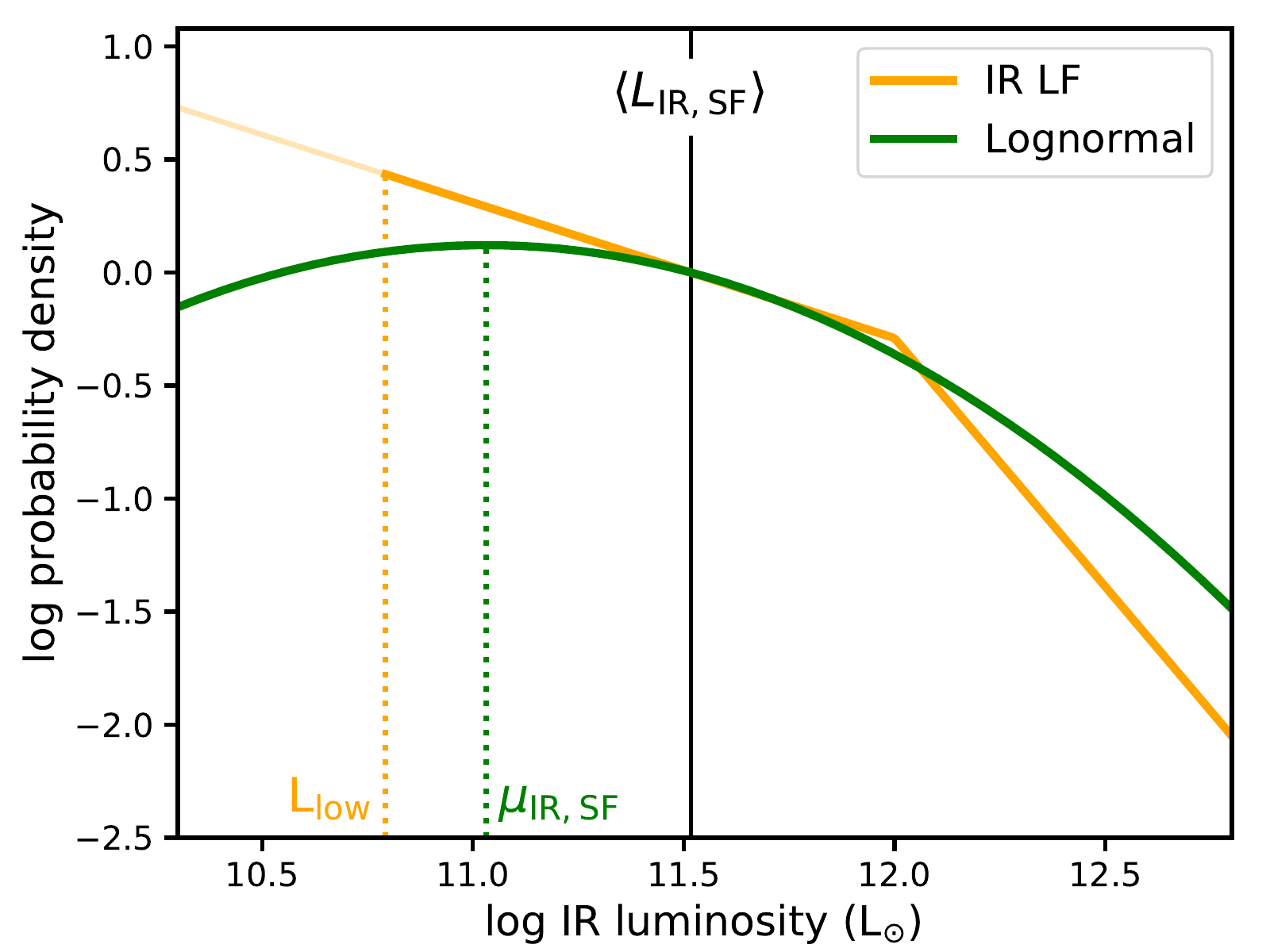}
\caption
{Illustration of the two forms of the IR luminosity distributions assumed in the Monte-Carlo simulations of
the radio emission from SF (Section \ref{sfsims} and Appendix \ref{irlfsims}). The lognormal distribution, plotted in green,
has a mode at $\mu_{\rm IR,SF}$ (the vertical dotted green line) and a redshift-independent width of w$_{\rm IR,SF}$ of 0.65 dex.
The distribution based on the IR luminosity function (IRLF), shown in orange, takes its overall shape from the IRLFs
in \citet{magnelli13}, with redshift-independent low and high luminosity slopes, and a low-luminosity cutoff ($L_{\rm low}$).
Both $\mu_{\rm IR,SF}$ and $L_{\rm low}$ are monotonically related to the average IR luminosity $\langle L_{\rm IR,SF} \rangle$,
shown by the solid vertical line, which is empirically determined for QSOs from the {\it Herschel}-based study of \citet{stanley17}.
Both distributions shown here share the same $\langle L_{\rm IR,SF} \rangle$, 
and their particular parameters are representative of cQSOs at $z\sim1.5$.
}  
\label{example_dists}
\end{figure}

The well-established infrared-radio correlation is a valuable tool to help us constrain the importance of strong starbursts
among rQSOs. Earlier studies with the {\it Herschel} and ALMA observatories have constrained the typical IR luminosities 
from SF-heated dust in QSOs. Using Monte-Carlo techniques, we can assess
the degree to which the observed radio emission of the LoTSS QSOs could arise from a star-forming population of QSO hosts 
with a distribution of IR luminosities motivated by these studies. 

We consider two approaches to describe the IR luminosity distributions of QSOs (Figure \ref{example_dists}). 
The first approach assumes that QSOs exhibit a lognormal distribution of IR luminosities arising
from SF. This assumption is motivated by ALMA continuum studies of distant AGN and QSOs \citep{scholtz18, schulze19}, as well
as theoretical considerations from hydrodynamic simulations \citep{scholtz18}. The second approach modifies
the redshift-dependent infrared luminosity functions (IRLFs) of galaxies to construct IR luminosity distributions tuned to QSOs. 
Both approaches yield qualitatively similar conclusions, and therefore we only present here our analysis based on the assumption of a
log-normal distribution, but refer the interested reader to the Appendix for the analysis using the IRLFs.

Our approach to connect the FIR distributions to the observed QSO population is as follows. 
%
Through an analysis of composite IR SEDs of SDSS QSOs from {\it Herschel} and {\it WISE} data, \citet{stanley17} 
measured the dependence of the mean IR luminosity attributable to SF as a function of 
redshift and AGN bolometric luminosity. In addition to a strong evolution with redshift, they 
reported a weak but definite trend in the average IR luminosities $\langle L_{\rm IR,SF} \rangle$ of 
QSOs with bolometric luminosity, which is governed primarily through a secondary correlation with SMBH mass \citep{rosario13, stanley17}.

Using \lsix\ and a 6 \mics\ bolometric correction $= 8$ \citep{richards06}, we constrain the median \lsix\ of the cQSOs
(i.e., typical SDSS QSOs)
and read off characteristic values for $\langle L_{\rm IR,SF} \rangle$ 
as a function of redshift directly from Figure 8 of \citet{stanley17}. 
Since we are reliant on their measurements, we perform our own analysis in bins of redshift that parallel
those used in that earlier work: $0.5<z<0.8$, $0.8<z<1.0$, $1.0<z<1.5$, $1.5<z<2.0$, $2.0<z<2.5$.
Following this approach, for cQSOs in these respective redshift intervals, 
we adopt $\langle L_{\rm IR,SF} \rangle$ of 44.6, 44.9, 45.1, 45.3, and 45.5 log \ergs.

$\langle L_{\rm IR,SF} \rangle$ is related monotonically to the mode of the lognormal function ($\mu_{\rm IR,SF}$)
that we assume is an adequate description of the IR luminosity distribution of the cQSOs:

\begin{equation}
\mu_{\rm IR,SF} \; = \; \log (\langle L_{\rm IR,SF} \rangle) \; - \; 1.15 \times w_{\rm IR,SF}
\end{equation}

\noindent where w$_{\rm IR,SF}$ is the logarithmic width of the distribution.
We adopted w$_{\rm IR,SF} = 0.65$ dex, fixed at all redshifts. This choice is motivated by the ALMA study 
of \citet{scholtz18} which constrained the empirical IR distributions of X-ray AGN in deep extragalactic fields. 
Very luminous QSOs at $z\sim2$ may show a narrower distribution
of SFRs \citep[width $\approx 0.5$;][]{schulze19}, but this level of difference in the assumed distribution width makes little 
qualitative difference to our results.

\begin{figure*}
\centering 
\includegraphics[width=\textwidth]{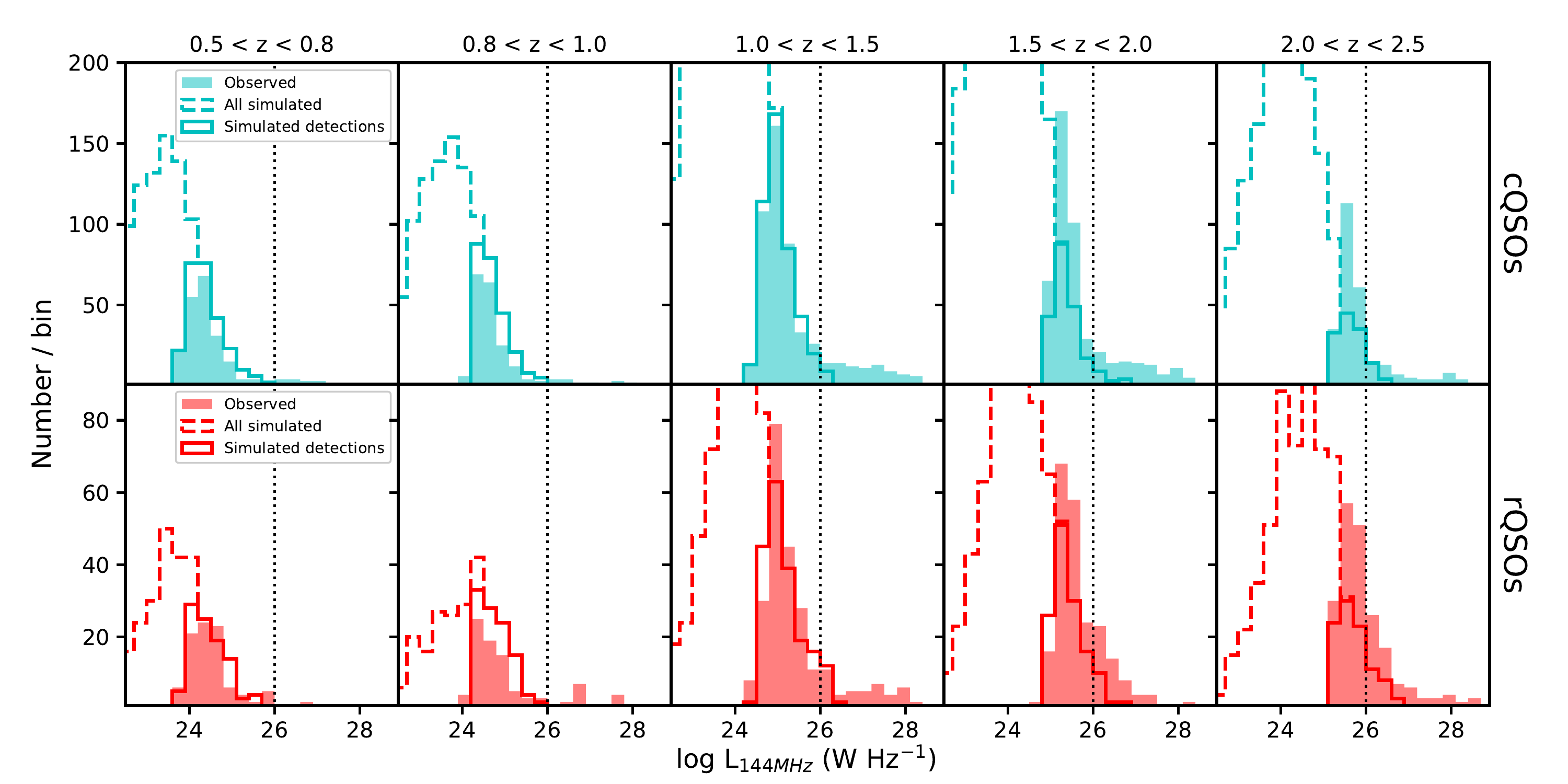}
\caption
{Distributions of radio luminosity (\lrad) comparing observed LoTSS-detected QSOs (filled histograms) and a simulation
of the radio emission from SF within the same QSOs (open histograms). The results for cQSOs (cyan)
are in the top panels and rQSOs (red) are in the bottom panels, while the populations are split into redshift bins
spanning left to right as indicated. The dashed-line open histograms show the simulated \lrad\ for all QSOs from the individual
colour-selected samples, while the solid-line open histograms only show the subset that would be ``detectable'' in LoTSS
at their redshifts and radio luminosities. The consistency between solid open and filled histograms indicate that SF
can be an important contribution to \lrad\ among the fainter end of the LoTSS radio QSOs. See Section \ref{sf_origins}
for more details. The dotted vertical black line marks a fixed luminosity of $10^{26}$ W Hz$^{-1}$.
}  
\label{lnorm_cr17}
\end{figure*}

For any given interval in redshift, we identify all cQSOs in the LoTSS DR1 field, irrespective of their radio properties.  
Then, treating the IR distribution function in that redshift interval as a probability distribution function, 
we randomly assign IR luminosities to all these cQSOs. 
Adopting the FIR-radio correlation appropriate for 144 MHz from \citet{calistro17}, including its redshift evolution and scatter, 
we are then able to convert these assigned IR luminosities to radio luminosities, and compare the predictions of this 
simulated distribution of \lrad\ to the actual \lrad\ distribution measured for the cQSOs.

In the top panels of Figure \ref{lnorm_cr17}, we plot the observed
\lrad\ distributions (filled histograms) and simulated \lrad\ distribution (dashed histograms) for cQSOs in the LoTSS DR1
field. To allow for the LoTSS detection limits, we additionally calculate observed-frame 144 MHz fluxes from
the simulated \lrad\ of the cQSOs, using their actual redshifts and a fixed $\alpha_{R} = -0.7$. 
In the Figure, we plot using solid open histograms the simulated luminosities for the subset of cQSOs
that are brighter than the nominal LoTSS DR1 5$\sigma$ detection limit of 0.35 mJy. 

An examination of top panels of Figure \ref{lnorm_cr17} indicates that, as per our nominal simulation, 
SF can account for as much as 50--100\% of the lower luminosity radio sources (\lrad$<10^{26}$ W Hz$^{-1}$) 
at all redshifts under consideration. In other words, most of the radio emission from the low-luminosity radio cQSOs can arise from
SF at a level consistent with the infrared luminosities of this population. Only the luminous tail of radio sources
are bright enough that SF cannot power their luminosities.

In the two highest redshift intervals, the nominal simulation does not fully recover the peak of the distribution of \lrad\
among radio-detected cQSOs (two right panels in the upper row of Figure \ref{lnorm_cr17}). This may mean that pure
SF is not enough to account for all the lower-luminosity radio sources among these QSOs. However, 
the simple application of an additional 0.2-0.3 dex to the nominal values of $\langle L_{\rm IR,SF} \rangle$ in these redshift intervals
is sufficient to reproduce the observed \lrad\ distributions. Such small differences are within the 
scope of the systematic uncertainties inherent to the SED fitting approach of \citet{stanley17}. 
We proceed with our analysis using the parameters of our nominal simulation, but note that such
uncertainties on the empirical IR luminosity distributions -- their averages and scatter -- can quantitatively
influence the comparisons of simulated and observed \lrad\ distributions. Therefore, we restrict
our conclusions to purely qualitative statements at this juncture, and refrain from making any quantitative measures
of statistical similarity between the distributions.
 
We obtain similar results\footnote{We have to increase $\langle L_{\rm IR,SF} \rangle$
by about 0.2 dex to reproduce the same relative numbers of LoTSS detections as our nominal simulation, 
but this is within the uncertainties of the measurements presented by \citet{stanley17}} 
if, instead of using the direct meter-wave FIR-radio correlation from \citet{calistro17}, 
we adopt the 1.4 GHz FIR-radio correlation from \citet{delhaize17} and a canonical $\alpha_{\rm R} = -0.7$. 
Since \citet{calistro17} and \citet{delhaize17} are independent studies that used different radio bands, different survey fields
and different analytic approaches, the similarity between the results of both simulations indicates a consistent treatment of
the SF contribution to the radio emission of the QSOs.

\subsubsection{Adapting the simulation to rQSOs} \label{rqsosims}

Turning now to the red QSOs, it is clear that the very same simulation parameters cannot fully
reproduce the increased fraction of LoTSS-detected rQSOs. 
In order to do so, there would have to be more rQSOs than cQSOs 
with simulated \lrad\ above the LoTSS detection threshold. Within the framework of the simulation,
we achieve this by ascribing a higher average SFR to the rQSOs, by increasing the mode $\mu_{\rm IR,SF}$
of the lognormal distribution that describes the IR luminosity of the rQSOs. After some experimentation, 
we found that a simple increase in $\mu_{\rm IR,SF}$ of 0.4 dex over the nominal value of the cQSOs is enough to account
for the redshift-averaged enhancement of the rQSO detection rate in LoTSS (a factor of $\approx 1.9$).
Implementing this increased SFR, we find that SF could also account for as much as 50--100\% of the low-luminosity 
rQSOs (lower panels of Figure \ref{lnorm_cr17}). 

LoTSS detects the steeply declining exponential tail of the star-forming QSO population, the starburst systems. 
Is there a maximum radio luminosity that SF can realistically power among the QSOs? 
In each panel of Figure \ref{lnorm_cr17}, we identify a radio luminosity of \lrad\ $ = 10^{26}$ W Hz$^{-1}$ as a vertical dotted line.
This luminosity is chosen as a benchmark for further analysis, because it marks a rough threshold beyond which
SF cannot account for the majority of the radio emission. From the simulation, less than 5\% of radio-detected 
rQSOs with \lrad\ $ > 10^{26}$ W Hz$^{-1}$ at any redshift can be powered by SF, even with the tweaks to the IR
luminosity distributions that are needed to reproduce the \lrad\ distributions of low-luminosity radio sources. Therefore, if 
it can be demonstrated that cQSOs and rQSOs more luminous than this threshold still show substantial differences, 
SF is unlikely to be root cause of the particular radio properties of red QSOs.

\subsubsection{Behaviour as a function of radio-loudness $\mathcal{R}$} \label{rlsims}

\begin{figure}
\centering 
\includegraphics[width=\columnwidth]{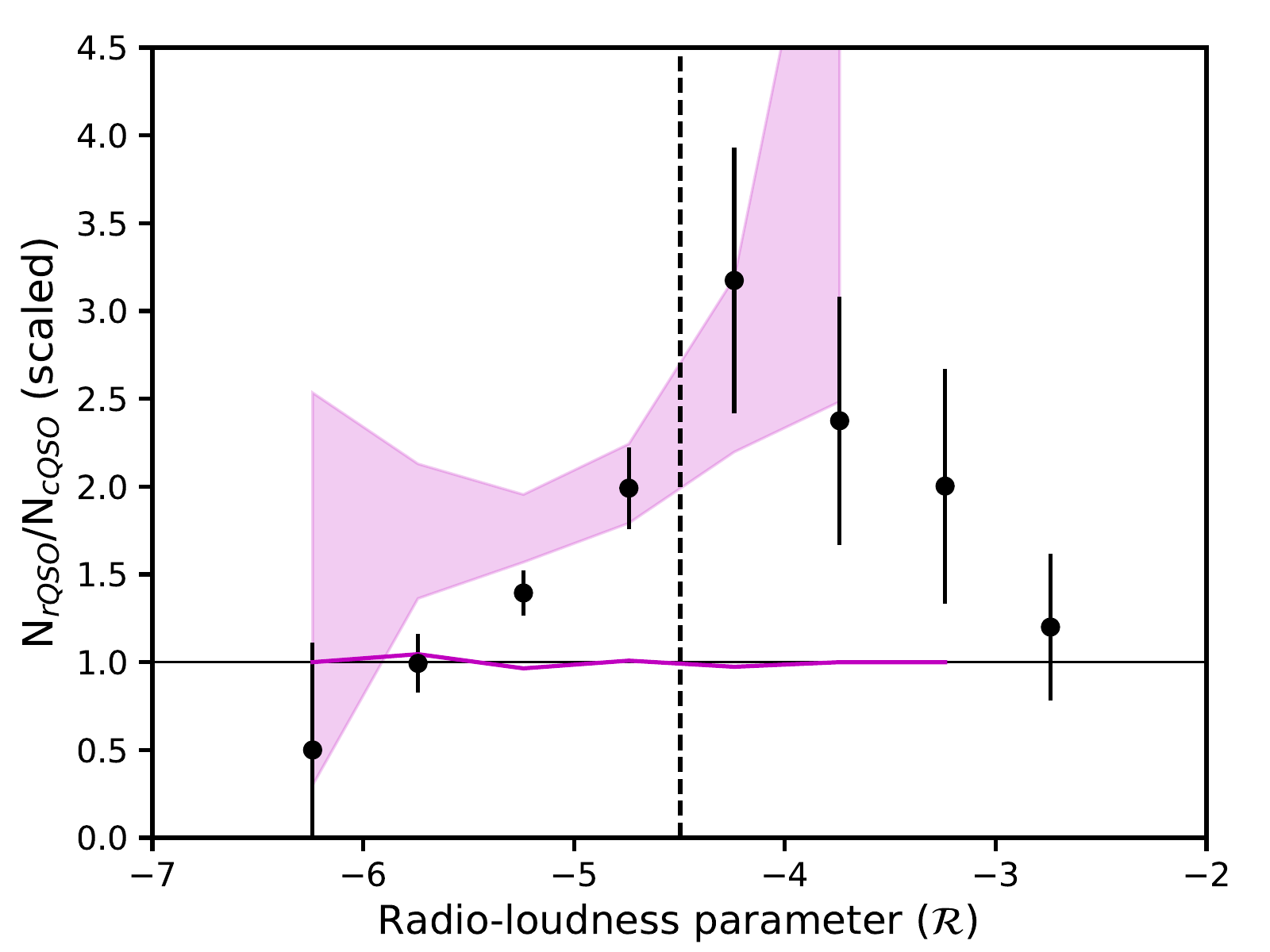}
\caption
{The ratio of the detection rate in LoTSS for rQSOs vs. L6$z$-matched cQSOs plotted in bins of the `radio-loudness'
$\mathcal{R}$, comparing the observed ratio (data points) against the predictions of our fiducial Monte Carlo simulation of SF in
QSOs (shaded magenta regions; see Section \ref{sf_origins} for details). 
These data are summarised from 30 bootstrap L6$z$-matching runs into the working sample sample in order
to overcome some of the sample variance inherent in the L6$z$-matching process.
The vertical dashed line at $\mathcal{R} = -4.5$ marks the threshold between systems that are
`radio-loud' (to the right) and `radio-quiet' (to the left). For reference, the solid magenta line shows
the median trend for a version of the simulation where rQSOs have the same IR distributions as the cQSOs.
}  
\label{detratio_R_simcomp}
\end{figure}

In our Monte-Carlo simulation, we probabilistically associated radio luminosities to the observed population of QSOs and
applied the flux limits of the LoTSS survey to determine the detectable fraction. We can go a step further and explore whether
the derived difference in the median SFRs of rQSOs over cQSOs (0.4 dex) can also reproduce the variation of the rQSO detection
enhancement with the radio-to-MIR luminosity ratio ($\mathcal{R}$; Section \ref{R_results} and Figure \ref{detratioR}). 

In Figure \ref{detratio_R_simcomp}, we show as a shaded region the predicted range of the radio detection enhancement
of rQSOs over L6$z$-matched cQSOs as a function of $\mathcal{R}$. As a baseline, we also show the median 
predictions of a simulation in which the rQSOs share the same IR luminosity distributions as the cQSOs (solid magenta line).
The observed enhancement is shown with the data points, in a similar fashion to Figure \ref{detratioR}. 
Both the simulated and observed data points plotted here are summarised from 30 independent bootstrap 
draws of an L6$z$-matched sample, which allows us to display a representative range of enhancements
shown by the L6$z$-matched QSOs. 

In comparing the shaded regions to the points, it must be kept in mind that the simulation does
not account for the luminous radio-loud sources which are undoubtedly powered by the AGN. Therefore,
while the observed data points extend beyond the radio-quiet/loud boundary (the vertical dashed line), the simulation
cannot reproduce these powerful systems. Our comparison is relevant only in the range of $\mathcal{R} = -6$ to $-4$,
where SF can make an important contribution to the detectable radio emission.

The simulation in which rQSOs have a increased modal SFR by 0.4 dex can broadly reproduce the enhancements
that we observe, and capture some of the dropoff towards low values of $\mathcal{R}$. At the lowest values of $\mathcal{R}<-5$,
the simulation does predict a higher enhancement than we see in the data. However, since the 
exact form of the dropoff is sensitive to details such as the assumed form of the IR luminosity distribution and 
the variation of $\langle L_{\rm IR,SF} \rangle$ with QSO bolometric luminosity, we only concentrate here on broad
trends rather than full quantitative agreement. It must also be remembered
that some of this consistency is because the SFR offset was \emph{tuned} to reproduce the observed redshift-averaged enhancement
in the detection fraction of rQSOs (Section \ref{rqsosims}). 

We conclude that the simulation is qualitatively successful at reproducing the behaviour of rQSOs
as a function of $\mathcal{R}$, but caution that this is not an independent test since it is sensitive to the tuning of the simulation
parameters.

\subsection{The origin of meter-wave radio emission in QSOs: accretion-powered emission} \label{agn_origins}

\begin{figure}
\centering 
\includegraphics[width=\columnwidth]{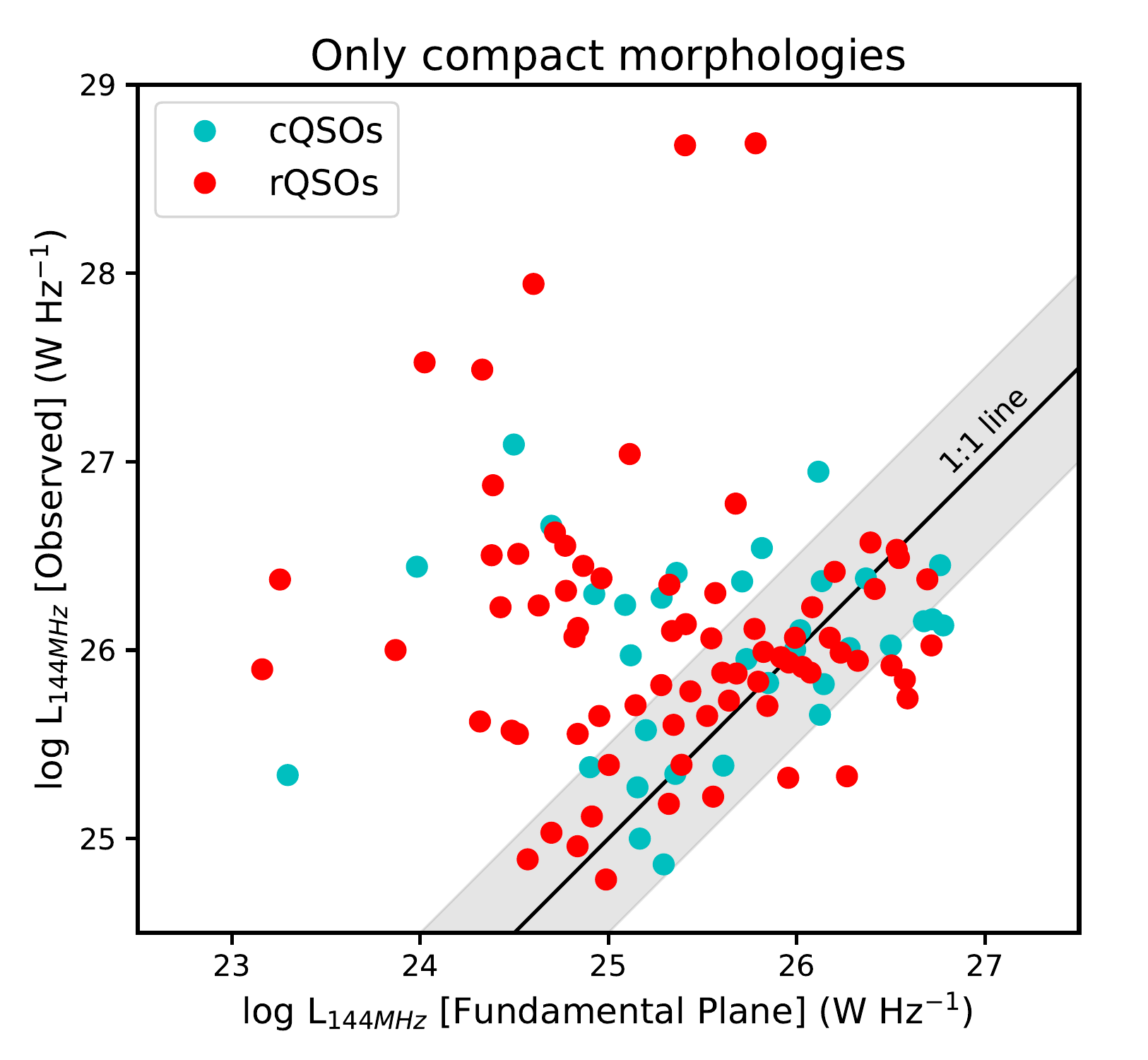}
\caption
{Observed 144 MHz radio luminosities (Y-axis) vs. predicted radio luminosities (X-axis) using the supermassive black hole ``Fundamental Plane''
relationship of \citet{merloni03}. Only QSOs with compact radio morphologies in LoTSS and black hole masses from the SDSS DR12
QSO catalogue \citep{kozlowski17} are are shown here. 
See Section \ref{agn_origins} for more details.
}  
\label{fp_compact}
\end{figure}

Our results indicate that the most substantial enhancement of the radio detection fraction 
in rQSOs is found among compact radio sources (Section \ref{radio_morphs}). We have examined SF
as a possible driver of the radio emission in such systems in the previous sub-section. Here we consider
whether the observed radio emission is consistent with an origin from AGN-related processes.

The most widely known mechanism of AGN-powered radio continuum emission 
is through bulk relativistic outflows (jets) ejected from the nucleus.
Another mechanism recently proposed as an important contribution in radio-quiet AGN is through particle acceleration in fast
shocks driven by thermal winds \citep[e.g.,][]{zakamska14}. Though the importance of the latter process is still under debate,
it is of particular interest in rQSOs because powerful obscured quasars are often associated with the kinematic signatures
of strong winds \citep{brusa15, zakamska16, klindt19}. In the analysis below, we use energetic arguments based on studies
of traditional jetted radio AGN, and briefly link in the expectations from the wind-shock scenario as well.

It is worth noting at this stage that strongly beamed synchrotron emission, such as that found in blazars or flat spectrum radio quasars,
cannot account for the rQSO population \citep{klindt19}. If such synchrotron emission was bright enough to contaminate the optical band
in our sources, their predicted radio luminosities would be orders of magnitude higher than we observe. 
Instead, among the most luminous radio QSOs we only find a low radio detection rate enhancement, in direct
contrast to the expectations based on pervasive beamed-synchrotron emission.

In the well-known empirical work of \citet{merloni03} which studied correlations in emission properties
across both stellar mass and SMBH accretion systems, the authors reported the discovery of a 
"Fundamental Plane" of black hole activity plane, a hypersurface in the three-dimensional space of
disc radiative emission (X-rays), core synchrotron emission (5 GHz) and black hole mass (M$_{BH}$).
While there is debate about the nature and origin of this Fundamental Plane, it serves a useful purpose in our study by providing
a reference level for the empirical radio emission expected from radiatively-efficient AGN.

A subset of the LoTSS DR1 radio QSOs have black hole masses estimated by \citet{kozlowski17}, as part of a value-added
spectral analysis of QSOs in the SDSS DR12 release. Concentrating only on the QSOs with compact radio morphologies,
which encompass all core-dominated radio sources, we are able to translate their \lsix\ and \lrad\ 
into the variables that specify the Fundamental Plane of \citet{merloni03}. To do this, we
assume a fixed ratio of 0.5 dex between the X-ray and 6 \mics\ luminosities of our QSOs \citep{lutz04} 
and take our usual approach of deriving or assuming a value of the radio spectral
index. This calculation gives us an expectation for the typical amount of radio emission these AGN should 
generate if they were translating their power into relativistic plasma.

In Figure \ref{fp_compact}, we plot the expected \lrad\ from the Fundamental Plane calculation against the measured
\lrad\ for radio QSOs from our working sample. 
A majority of both cQSOs and rQSOs with compact radio emission have radio luminosities that are 
comparable to or more powerful than the predictions of the fundamental plane. 
As the Fundamental Plane is only applicable to the nuclear or core radio emission, we interpret
the scatter of QSOs with higher observed \lrad\ as likely arising from extended jetted emission that is unresolved
by LoTSS at the 6'' angular resolution of the survey. 
From the analysis summarised in Figure \ref{fp_compact}, we can conclude that the radio luminosities of the QSOs 
are at least broadly consistent with an origin from AGN.

In Section \ref{R_results}, we identified a peak in the radio detection enhancement among rQSOs at a preferred 
value of the radio-loudness parameter $\mathcal{R} \approx -4$. Under the hypothesis that the radio emission in QSOs
is powered by an AGN at these intermediate values of radio-loudness, it is useful to case 
the characteristic value of $\mathcal{R}$ in more physical terms. The radio luminosity of a relativistic outflow
is believed to be connected to its total kinetic power, though the exact relationship is only known in dense cluster environments
where both quantities can be independently measured. Nevertheless, if we adopt such a relationship to calculate the kinetic
power of a radio source \citep[e.g., Equation 6 from][]{merloni07}, and take a standard 
6 \mics\ bolometric correction for QSOs \citep[$\approx 8$,][]{richards06}, we can convert $\mathcal{R}$ to a ratio of 
kinetic-to-bolometric power in a AGN-powered radio source as follows:

\begin{equation}
\log \frac{\rm{Kinetic}}{\rm{Bolometric}} \; \approx \; 0.8\, \mathcal{R} - 0.2\, L_{\rm 6 \mu m} + 11.5
\label{eqn2}
\end{equation}

\noindent The mild dependence of \lsix\ is because of the sub-linear
relationship derived by \citet{merloni07} between the kinetic and synchrotron power of a radio source.

The wind-shock scenario of \citet{zakamska14} assumes a conversion efficiency between the wind kinetic energy and synchrotron radio power
that is equivalent to starburst superwinds, a typical value of $\approx 4.5$ dex. This is quite close to the ratio in relativistic
jetted outflows reported by \citet{merloni07}. Therefore, our inferences are not very sensitive to the driving mechanism
of the synchrotron emission, certainly within the large systematic uncertainties that go into such calculations.

From Equation \ref{eqn2}, we estimate that a QSO with $\log \rm{L}_{6 \mu m} = 44.5$--$46.5$ \ergs\ lying on the $\mathcal{R} = -4.5$
division puts out about 3-10\% of its power in kinetic form. Interestingly, this is approximately the range expected for the
efficiency of AGN feedback in successful simulations of galaxy evolution \citep[e.g.][]{bower06,angles17}.

\subsection{What is responsible for the distinct radio properties of red QSOs?} \label{synthesis}

Our simulation of the radio luminosities attributable to SF in QSO hosts indicates that a substantial proportion of the
lower-luminosity radio sources detected by LoTSS, including a majority of those that lie below the FIRST 
detection limit, could be powered by SF. On the other hand, the luminosities of the compact radio
sources, the category most over-represented among rQSOs, are quite consistent with an AGN origin.
In this subsection, we bring together various lines of evidence laid out in this work to explore whether 
the distinguishing properties of rQSOs favour an enhanced level of SF or arise instead 
from a special AGN-powered population found mostly among rQSOs.

From the simulation in Section \ref{sf_origins}, 
we concluded that a difference in the median amount of SF at the level of 0.4 dex is enough
to distinguish the \lrad\ distributions of rQSOs and cQSOs in LoTSS. The SF in these QSOs will, in the most part,
be found only on scales of their host galaxies, i.e, a few -- 10 kpc. Therefore, we expect the star-forming
radio emission to be unresolved by LoTSS, consistent with our result that the enhanced rQSO detection rate 
is only found among compact radio sources, in which the central radio emission is the dominant component of the total intensity.

Meter-wave radio emission from SF is primarily optically-thin synchrotron emission from supernova remnants and
shows a very similar range of meter-wave spectral indices as AGN radio jets. 
Therefore, we would not expect strong differences in the spectral indices of 
SF-dominated radio QSOs compared to those that may be dominated by compact radio jets \citep[e.g.,][]{calistro17}. 
This is consistent with the similar spectral indices as a function of morphology that we find in our sample (Figure \ref{morph_alphadists}).

However, a closer look at the \lrad\ distributions may imply a different interpretation.
A valuable bit of insight from the simulations of Section \ref{sf_origins} is that radio
sources with \lrad$ > 10^{26}$ W Hz$^{-1}$ are very unlikely to arise from SF. 
Yet, from Figure \ref{morph_lrdists}, approximately half of the colour-selected
QSOs with compact radio sources have radio luminosities above this threshold. If SF was the only driver
for the different radio properties of the rQSOs, the more luminous compact radio sources, which cannot be dominated by SF,
should be found at roughly the same rate of incidence in cQSOs and rQSOs.

In the right panel of Figure \ref{vismorphs}, we compare the distribution of morphologies for colour-selected QSOs with
\lrad$ > 10^{26}$ W Hz$^{-1}$. Since this restricts our analysis only to the brighter LoTSS radio QSOs, the fraction of 
`Faint' sources drops drastically as expected, while the sources in the other morphological categories suffer
smaller reductions in their fractions because they are generally brighter. 
The main point to note from this Figure is that, despite the application of our \lrad\ threshold,
the relative enhancement in the fraction of rQSOs over cQSOs remains about the same in both the `Compact' and the 
`Marginally Extended' categories. Therefore, the radio luminosities of the 
special population of compact radio-quiet sources found among rQSOs
extends beyond the range where SF can be reasonably
expected to explain the differences. This supports the notion that the primary driver of the enhancements seen among
rQSOs is from a population of compact, AGN-powered radio sources.

In Figure \ref{detratioR}, we examined the observed variation of 
the radio detection enhancement in rQSOs vs. the radio-loudness parameter $\mathcal{R}$, finding
a more pronounced peak in the enhancement at $\mathcal{R} \sim -4$, with a strong drop off towards
lower values of $\mathcal{R}$. This suggests that the special population of rQSOs responsible for the enhancement 
occupies a preferred radio-loudness, slightly below the canonical boundary between radio-loud and radio-quiet systems.
How do we interpret this result in terms of the interplay between SF and AGN-powered radio emission in QSOs?

Our analysis is Figure \ref{detratio_R_simcomp} shows that a scenario where rQSOs have the same level of AGN-powered
radio emission but a higher level of SF (by 0.4 dex) can qualitatively explain the trend of increasing rQSO detection
enhancement with $\mathcal{R}$ seen among radio-quiet QSOs. On the other hand, from our 
morphological assessment (Figure \ref{vismorphs} and discussion above), 
we find a substantial detection enhancement even among radio sources which are luminous enough that they
cannot be reasonably powered by SF. 

We can reconcile these conflicting points of evidence by considering an alternative explanation for the trends with $\mathcal{R}$:
the differences between rQSOs and cQSOs arise from an underlying population of radio-intermediate AGN-powered 
radio sources that are over-represented among rQSOs. In this scenario, both colour-selected QSO subpopulations have similar 
SFR distributions, and the dropoff that we see towards low values of $\mathcal{R}$ in Figure \ref{detratioR} is because
very radio-quiet systems, whether red or blue, are increasingly dominated by SF.

Adequately distinguishing between these two explanations requires a careful assessment of the level of SF in 
rQSOs {\it vis-\`a-vis} cQSOs. In the accompanying study of \citet{fawcett20}, we have used multi-wavelength constraints 
to identify SF-dominated SDSS QSOs within the 2 deg$^{2}$ COSMOS extragalactic deep field.
We find that such systems are invariably very radio-quiet ($\mathcal{R} < -5$), and that AGN-dominated radio-quiet
QSOs potentially still show an enhancement in their radio detection rates. This bolsters the argument that differences
in SF are not the root cause of the enhanced radio emission among rQSOs.

On-going radio studies at high angular resolution, designed to resolve the kpc-scale emission in 
rQSOs and cQSOs, may be able to shed more light on this phenomenon. In addition, we are undertaking a deep rest-frame 
far-IR imaging study of SDSS colour-selected QSO with the Atacama Large Millimetre Array (ALMA) which will help us directly measure
the IR-based SFRs in rQSOs and cQSOs, and conclusively determine whether red QSOs are associated
with enhanced SF.

Regardless of the real situation, our results require that either the distributions of SFR or AGN-driven 
radio emission in red QSOs are fundamentally different from typical QSOs. This implies that 
environments of these systems, whether on nuclear, host galaxy, or larger scales, are also fundamentally different.
A thorough exploration of these differences is essential in understanding the role of the red QSO phase
in the co-evolution of galaxies and supermassive black holes.

\section{Conclusions}

Using the SDSS DR14 QSO catalog and the LoTSS DR1 survey, we compare the low-frequency radio properties
of red QSOs (rQSOs) and typical QSOs (cQSOs). We use {\it WISE}-based MIR photometry and a statistical matching technique
to control for differences in redshift and intrinsic luminosities between the two subsamples. We find:

\begin{enumerate}

\item The population of rQSOs are $\times3$ more likely to be detected in LoTSS than equivalent cQSOs (Section \ref{det_stats}), 
consistent with an earlier FIRST-based study from our team \citep{klindt19}.  

\item The sources that drive the higher radio detection fractions of rQSOs have radio morphologies that are primarily
compact (Section \ref{radio_morphs}), further confirming the findings of \citet{klindt19}. The radio luminosities and 
spectral indices of the compact rQSOs are indistinguishable from cQSOs (Section \ref{radio_alphas})

\item Much of the difference between rQSOs and cQSOs can be attributed to an excess population of radio-intermediate sources, with
a radio-loudness parameter $\mathcal{R} \approx -4$ (Section \ref{R_results}, where $\mathcal{R}$ is a measure of the MIR-to-radio power). 

\item By statistically modelling the expected amount of star formation (SF) in the host galaxies of QSOs, we demonstrate that SF
is a major source of radio emission in the radio-quiet regime in both rQSOs and cQSOs (Section \ref{sf_origins})

\item Bringing together various lines of evidence, including the characteristic radio luminosities at which rQSOs show the most pronounced
radio excess, we conclude that their special properties likely arise from processes that are linked
to their AGN, rather than due to higher levels of SF (Section \ref{synthesis}). Our results 
confirm, quite spectacularly, that dust-reddened QSOs are a fundamentally different population of AGN than the more common blue QSOs.
  
 \end{enumerate}

\section*{Acknowledgements}

We thank the anonymous referee for their valuable input in the final stages of this work.
D.R. and D.M.A. acknowledge the support of the UK Science and Technology Facilities Council (STFC) through grants 
ST/P000541/1 and ST/T000244/1. L.K. acknowledges the support of a Faculty of Science Durham Doctoral Scholarship.
Funding for the SDSS has been provided by the Alfred P. Sloan Foundation, the U.S. Department of Energy Office of Science, and the Participating Institutions. SDSS-IV acknowledges support and resources from the Center for High-Performance Computing at
the University of Utah. The SDSS web site is \href{http://www.sdss.org}{www.sdss.org}. SDSS is managed by the Astrophysical Research Consortium for the Participating Institutions of the SDSS Collaboration. We make use of data products from the Wide-field Infrared Survey Explorer, which is a joint project of the University of California, Los Angeles, and the Jet Propulsion Laboratory/California Institute of Technology, funded by the National Aeronautics and Space Administration.

\bibliographystyle{mn2e}

\bibliography{rqso_lotss}

\appendix

\section{Simulating the SFR contribution in typical QSOs: the IRLF approach} \label{irlfsims}

\begin{figure*}
\centering 
\includegraphics[width=\textwidth]{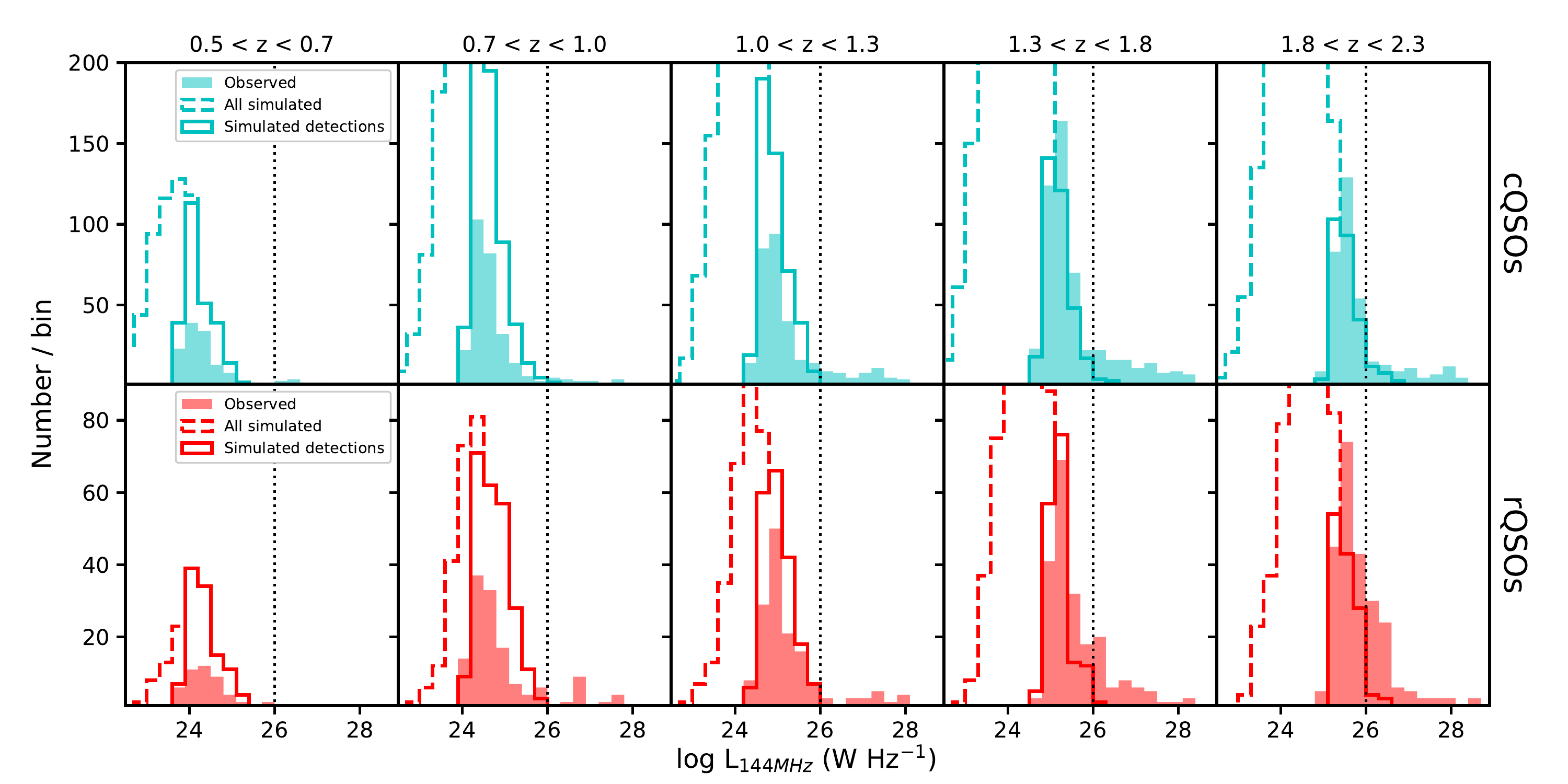}
\caption
{Distributions of radio luminosity (\lrad) comparing observed LoTSS-detected QSOs (filled histograms) and a simulation
of the radio emission from SF within the same QSOs (open histograms). This version of the simulation
assumes a distribution of IR luminosities in QSOs that follows the IR luminosity function of galaxies with a variable
low-luminosity cutoff (see Section \ref{irlfsims} for details).
The results for cQSOs (cyan)
are in the top panels and rQSOs (red) are in the bottom panels, while the populations are split into redshift bins
spanning left to right as indicated. The dashed-line open histograms show the simulated \lrad\ for all QSOs from the individual
colour-selected samples, while the solid-line open histograms only show the subset that would be ``detectable'' in LoTSS
at their redshifts and radio luminosities. The dotted vertical black line marks a fixed luminosity of $10^{26}$ W Hz$^{-1}$.}  
\label{irlf_cr17}
\end{figure*}

In Section \ref{sfsims}, we described our approach of simulating the radio emission from SF in QSOs.
In our primary simulations, we assumed that the IR luminosity distributions of SF heated dust in 
QSOs could be adequately described with lognormal functions (Figure \ref{example_dists}). Alternately, one could instead
assume that QSO host galaxies share the shape of the IR luminosity function (IRLF) of the co-eval galaxy population, and share
in the general evolution of the IRLF. Here we consider a simulation based on the galaxy IRLF, and apply it to the
SDSS DR14 cQSOs and rQSOs in the LoTSS DR1 region.

In the last decade, several studies with the FIR {\it Herschel} Space Observatory have accurately constrained 
the redshift-dependent IRLFs of galaxies over areas of sky comparable to the LoTSS HETDEX field (hundreds of deg$^{2}$).
These IRLFs empirically capture the space density of the luminous tail of the infrared galaxy population,
which are responsible for most of the SF dominated systems detectable in LoTSS beyond the local Universe 
(Figure \ref{radio_ztrends}). For the purposes of our calculations, we use the IRLFs from \citet{magnelli13}, a composite study that combined 
{\it Herschel}  constraints from both large-area and deep fields.
These IRLFs are described with a simple broken power-law function, with fixed low-and high-luminosity slopes of $-0.6$
and $-2.2$ that connect at a characteristic break luminosity ($L^{\rm IR}_{\star}$). Since we treat these IRLFs are probability
distribution functions that integrate to unity, we do not use their absolute normalisations in our analysis.

The integral of the IRLF described above is unbounded if the lower integration limit $\to -\infty$. Therefore, in order
to use it as a distribution function, we impose a finite lower limit to the integral, which functionally acts as a low luminosity
cutoff on the IRLF ($L_{\rm low}$). For a given $L^{\rm IR}_{\star}$, $L_{\rm low}$ is monotonically related to the average
IR luminosity of the modified IRLF $\langle L_{\rm IR,SF} \rangle$ (see Figure \ref{example_dists} for a visualisation).
\citet{stanley17} have empirically determined $\langle L_{\rm IR,SF} \rangle$ for SDSS QSOs using {\it Herschel} and {\it WISE} 
stacked photometry. We use their relations to set $L_{\rm low}$ as a function of redshift and the median bolometric luminosities of 
given QSO subpopulation.

For the IRLF simulation, we follow an approach similar to our primary simulation to 
estimate the 144 MHz radio luminosity (\lrad) distributions of QSOs. We refer
the reader to Section \ref{sfsims} for the essential details. 
The key differences to the aforementioned approach is (a) that we use a different family of IR luminosity distribution functions, and (b)
that we adopt a different redshift binning scheme, to remain consistent with the binning used by \citet{magnelli13}
for their determination of the IRLF and its evolution: $0.5<z<0.7$, $0.7<z<1.0$, $1.0<z<1.3$, $1.3<z<1.8$, and $1.8<z<2.3$.

In Figure \ref{irlf_cr17}, we compare the distributions of \lrad\ predicted by IRLF simulation, for the full sample and the
`detectable' subset, against the observed \lrad\ distributions of cQSOs and rQSOs from the working sample. In this version
of the simulation, we achieve the higher incidence of starbursts required among rQSOs by increasing the cutoff luminosity 
$L_{\rm low}$ in the rQSO simulation by 0.4 dex, which reproduces the higher redshift-averaged LoTSS detection 
fraction seen in this subset of colour-selected QSOs.

A direct comparison of Figure \ref{irlf_cr17} and Figure \ref{lnorm_cr17} indicates that 
both simulations yield the same basic qualitative
conclusion that most, if not all, of the lower luminosity radio QSOs in LoTSS (\lrad\ $< 10^{26}$ W Hz$^{-1}$) can be powered by SF.
They agree in this regard despite the different underlying IR luminosity distributions that we assumed.
However, there are some important differences. The simulation based on the IRLF overestimates the number of detectable
QSOs at $z<1.3$ by factors of a few. This can be significantly mitigated by reducing the values of 
$\langle L_{\rm IR,SF} \rangle$ used in this simulation by $0.3$-$0.4$ dex. However, these changes to the mean IR luminosity
of the QSOs would make them inconsistent with the empirical measurements of \citet{stanley17}. 

Another, probably more realistic, 
way to bring the simulated and observed distributions in better agreement is to by allow a more gradual low luminosity
drop off in the shape of the modified IRLF, in place of a sharp cutoff at $L_{\rm low}$. However, as one may infer from
Figure \ref{example_dists}, this leads to a functional form that resembles the lognormal function adopted in the
standard simulations of Section \ref{sfsims}. For this reason, we do not pursue the IRLF-based simulations for any
further scientific analysis beyond the simple confirmation of our inferences regarding the importance of SF in powering 
low-luminosity radio QSOs in LoTSS.

\end{document}